%% file: ms.tex
\documentclass[a4paper,11pt]{article}
\usepackage[mathlines]{lineno}

\usepackage{jcappub}

\usepackage[table,svgnames,dvipsnames]{xcolor}
\usepackage[normalem]{ulem}
\usepackage{aas_macros}
\usepackage{hyperref}
\usepackage{verbatim}
\usepackage{tabularx}
\usepackage{orcidlink} % for author list
\usepackage{hanging} % for affiliations 
\usepackage{multirow}
\usepackage{makecell}
\usepackage{siunitx}
\usepackage{dcolumn} % Align table columns on decimal point
\usepackage{bm} % bold math
\usepackage{xspace}
\usepackage{graphicx} % Include figure files
\usepackage{subfigure}
\usepackage{cases} % for dealing with mathematics,
\usepackage{booktabs}

\usepackage[nameinlink,noabbrev]{cleveref}
\crefname{equation}{Eq.}{Eqs.}
\crefname{section}{Section}{Sections}
\crefname{figure}{Figure}{Figures}
\crefname{table}{Table}{Tables}
\crefname{appendix}{Appendix}{Appendices}
\Crefname{figure}{Figure}{Figures}
\Crefname{equation}{Equation}{Equations}
\Crefname{section}{Section}{Sections}
\Crefname{table}{Table}{Tables}

\newcommand{\alperp}{\alpha_{\perp}}
\newcommand{\alpara}{\alpha_{\|}}
\newcommand{\aiso}{\alpha_\mathrm{iso}}
\newcommand{\aAP}{\alpha_\mathrm{AP}}
\newcommand{\elgz}{\texttt{ELGO}\,}
\newcommand{\elgo}{\texttt{ELG1}\,}
\newcommand{\elgt}{\texttt{ELG2}\,}
\newcommand{\lrgo}{\texttt{LRG1}\,}
\newcommand{\lrgt}{\texttt{LRG2}\,}
\newcommand{\lrgth}{\texttt{LRG3}\,}
\newcommand{\bgs}{\texttt{BGS}\,}
\newcommand{\qso}{\texttt{QSO}\,}

\newcommand{\mpch}{h^{-1}\,\mathrm{Mpc}}
\newcommand{\hmpc}{h\,\mathrm{Mpc}^{-1}}
\newcommand{\xiellprime}{\xi^\prime_\ell(s)}
\newcommand{\xiell}{\xi_\ell(s)}

\def\2pr{^{\prime \prime}}
\def\deg{^{\circ}}

\def\geqsim{\lower.73ex\hbox{$\sim$}\llap{\raise.4ex\hbox{$>$}}$\,$}
\def\leqsim{\lower.73ex\hbox{$\sim$}\llap{\raise.4ex\hbox{$<$}}$\,$}

\newcommand{\ebv}{$E(B-V)$ }
\newcommand{\sysnet}{\textsc{SYSNet} }
\newcommand{\obiwan}{\textsc{Obiwan} }
\newcommand{\regressis}{\textsc{Regressis} }
\newcommand{\obiwans}{\textsc{Obiwan}'s }
\newcommand{\obisys}{\textsc{Obiwan}+\textsc{SYSNet} }

\title{Mitigating Imaging Systematics for DESI 2024 Emission Line Galaxies and Beyond}

\input{authorlist}

\abstract{Emission Line Galaxies (ELGs) are one of the main tracers that the Dark Energy Spectroscopic Instrument (DESI) uses to probe the universe. However, they are afflicted by strong spurious correlations between target density and observing conditions known as imaging systematics. 
In this paper, we present the imaging systematics mitigation applied to the DESI Data Release 1 (DR1) large-scale structure catalogs used in the DESI 2024 cosmological analyses. We also explore extensions of the fiducial treatment. This includes a combined approach, through forward image simulations (\textsc{Obiwan}) in conjunction with neural network-based regression, to obtain an angular selection function that mitigates the imaging systematics observed in the DESI DR1 ELGs target density. 
We further derive a line of sight selection function from the forward model that removes the strong redshift dependence between imaging systematics and low redshift ELGs. Combining both angular and redshift-dependent systematics, we construct a three-dimensional selection function and assess the impact of all selection functions on clustering statistics. We quantify differences between these extended treatments and the fiducial treatment in terms of the measured 2-point statistics. We find that the results are generally consistent with the fiducial treatment and conclude that the differences are far less than the imaging systematics uncertainty included in DESI 2024 full-shape measurements.
We extend our investigation to the ELGs at $0.6<z<0.8$, i.e., beyond the redshift range ($0.8<z<1.6$) adopted for the DESI clustering catalog, and demonstrate that determining the full three-dimensional selection function is necessary in this redshift range. 
Our tests showed that all changes are consistent with statistical noise for BAO analyses indicating they are robust to even severe imaging systematics. Specific tests for the full-shape analysis will be presented in a companion paper.}

\begin{document}
\maketitle
\flushbottom

% BODY OF PAPER 

\section{Introduction}
\label{sec:introduction}
Understanding the nature of dark energy is of great importance to modern cosmology and is the primary goal of the 
Dark Energy Spectroscopic Instrument (DESI) \cite{Snowmass2013.Levi,DESI2016a.Science,DESI2016b.Instr,DESI2022.KP1.Instr,FocalPlane.Silber.2023,Corrector.Miller.2023,DESI2023a.KP1.SV,DESI2023b.KP1.EDR,Spectro.Pipeline.Guy.2023} Stage-IV \cite{albrecht2006report} experiment. DESI is a multi-fiber spectroscopic instrument installed on the Mayall 4-m telescope that, through carefully planned observations \cite{SurveyOps.Schlafly.2023}, efficiently measures galaxy redshifts \cite{Anand.RedshiftEstimation.2024,Redrock.Bailey.2024} with the purpose of creating a 3D map of the Universe \cite{Snowmass2013.Levi}.
From this LSS mapping of the Universe, the DESI collaboration generates summary statistics from galaxy clustering and probes dark energy through the measurement of baryon acoustic oscillation (BAO) and redshift space distortions (RSD). BAO \cite{DESI2024.III.KP4, DESI2024.IV.KP6,DESI2024.VI.KP7A} and RSD \cite{DESI2024.V.KP5,DESI2024.VII.KP7B} measurements will inform us about the expansion history of the Universe and the growth of structure, respectively. Measurements of the clustering at the largest scales constrain the amount of Primordial non-Gaussianity (PNG) \cite{ChaussidonY1fnl}.

However, the selection of the tracers used in spectroscopic surveys such as DESI relies on photometry from imaging surveys, and this photometry is not perfect. We know that observing conditions and foreground maps, also called imaging systematics, can present non-trivial correlations with the tracers. These complex correlations can then impart spurious fluctuations in the observed density field, which would make the observed galaxy clustering differ from the truth. Then, as we could expect, tracers sensitive to imaging systematics will propagate uncertainties into our cosmological measurements. Hence, efficient and careful methods to handle this source of contamination must be developed to obtain robust cosmological constraints.

The BAO measurement historically has been proven to be robust against imaging systematics, due to its characteristic feature in the 2-pt correlation function \cite{ross_2012,RossDR12,eBOSS_BAO-RSD_2020}. Conversely, constraining PNG has been proven to be especially sensitive to imaging systematics, as shown in previous works \cite{Rezaie_2021,rezaie2023local} and others since its constraints strongly depend on large scales, low wavenumber ($k$) or high separation ($s$), at which scales the intrinsic galaxy clustering amplitude can become dominated by variations in observing conditions. 
RSD fits are part of a full-shape analysis of the power spectrum, also potentially making it sensitive to imaging systematics since they can dramatically alter the observed amplitude. The impact of image systematics on DESI 2024 full shape is discussed in \cite{KP5s6-Zhao}.

The effect of imaging systematics in cosmological analysis have been thoroughly studied across various survey programs: SDSS \cite{Scranton.SDSS.2002,Myers.SDSS.2006,ross_2011,Ho.SDSS.2012,Pullen_2013,Leistedt.SDSS.2013}, DES \cite{DES2016,Elsner2016,DESY1.2018,DESY3.2021}, BOSS \cite{Ross_2016,Ross.eBOSS.2020}, eBOSS \cite{Laurent_2017}.
Different approaches have been explored when trying to ameliorate the effect of imaging systematics of these surveys. 
Among most methods of systematic treatment, it is assumed that the correlation between systematic properties and the tracer density field is linear. The mitigation schemes that rely on this assumption include linear regression, mode projection and template subtraction. In \cite{Weaverdyck.compsys.2021} they review these mitigation methods and demonstrate how the pseudo-power spectrum mode projection method can be interpreted as a linear regression.
Nevertheless, these mitigation schemes are not sufficient to completely remove large-scale contamination of strongly contaminated samples, since they assume a trivial relationship between imaging systematics and observed target density.  

More recently developed approaches consider image simulations or the use of deep learning. Developed for DESI imaging, \obiwan \cite{obiwan, Kong_2020} forward models with Monte Carlo image simulations that inject fake sources in imaging data and simulate how the pipeline generates the survey catalog models and selects its targets.
For DES-Y3 they also forward-model the survey selection function through the \textsc{Balrog} image simulation \cite{DESY3.2022}.
On the deep-learning side,  \cite{Rezaie_2020, Rezaie_2021,rezaie2023local} implements a fully connected feed-forward Neural Network (NN) that learns the non-linear correlations between imaging templates and target density. Also, in \cite{QSO.TS.Chaussidon.2023} they implement a Random Forest algorithm to handle the angular selection of DESI quasars.

DESI target selection is based on imaging obtained from the 9th Data Release of the Legacy Survey \cite{LS.Overview.Dey.2019,LS.dr9.Schegel.2024}, which itself is composed of several observational programs. The programs that provided photometry are the Beijing-Arizona Sky Survey (BASS, \cite{BASS.Zou.2017}), the Mayall z-band Legacy Survey (MzLS, \cite{MzLS}), the Dark Energy Camera Legacy Survey (DECaLS, \cite{LS.Overview.Dey.2019}), and the Dark Energy Survey (DES, \cite{DES}). Each of these imaging surveys has imprints left by imaging systematics, due to observing conditions, that could bias the true cosmological signal.

This is particularly true for DESI ELGs. With the DESI instrument we can efficiently observe ELGs up to $z=1.6$, obtaining secure redshifts with the [O II] doublet feature in the ELG spectra. However, the DESI ELG target sample is heavily contaminated due to its faint imaging flux limit and is strongly sensitive to galactic extinction and galactic depth \cite{elg_selection}. Since DESI ELGs suffer from such contamination, methods must be developed to carefully treat the sample and understand its systematic variation, so that any cosmological inference is not biased. Our work focuses on imaging systematics; the impact of spectroscopic systematics on DESI ELG clustering is studied in \cite{KP3s3-Krolewski,KP3s4-Yu}.

This work is outlined as follows. 
In Section \ref{sec:data}, we describe the data and imaging maps used in our analysis.
Section \ref{sec:analysis_methods} details the methods we use to measure 1-pt and 2-pt statistics in our analysis.
In section \ref{sec:sysnet}, we present the systematic mitigation method developed and applied to DESI Data Release 1 (DR1; \cite{DESI2024.I.DR1}) ELGs LSS catalogs \cite{DESI2024.II.KP3}, which were used for DESI 2024 cosmological analyses.
In Section \ref{sec:obiwan} we outline the forward image simulation and its results.
In Section \ref{sec:dndz}, we study the correlation between imaging properties and variation in the radial (redshift) distribution of the ELG sample and develop a method to mitigate such variation.
In Section \ref{sec:obinn}, we introduce and test a new approach to systematic mitigation applied to the same DESI DR1 ELG sample, which combines forward image simulations and network-based regression.
In Section \ref{sec:clustering_stats}, we quantify the impact of imaging systematics on 2-pt clustering statistics, i.e. the correlation function and its Fourier counterpart, the power spectrum. We also test the robustness of the BAO measurement against systematic treatment for all DESI 2024 tracers. 
Finally, in Section \ref{sec:conclusion} we summarize our findings and suggestions for future studies.

\section{Data}
\label{sec:data}
DESI target selection uses imaging data in different optical bands from Data Release 9 of the Legacy Survey (DR9-LS). The DR9-LS photometry is composed of several survey programs that observed in various optical bands and different regions of the sky. For imaging systematic analysis we are concerned about the North (above declination of $32.375\deg$) and the South (below declination of $32.375\deg$) regions of the sky, since they possess distinct photometric properties. The North is composed of photometry from the North Galactic Cap (NGC), where its $g$ and $r$ band photometry is provided by the Beijing-Arizona Sky Survey (BASS, \cite{BASS.Zou.2017}) and its $z$ band from the Mayall $z$-band Legacy Survey (MzLS, \cite{MzLS}).  The South contains photometry from both NGC and the South Galactic Cap (SGC). All South data was obtained using the Dark Energy Camera (DECam, \cite{DECam}) which collected measurements in the $g$, $r$ and $z$ bands for the Dark Energy Camera Legacy Survey (DECaLS, \cite{LS.Overview.Dey.2019} provided for NGC and SGC) and the Dark Energy Survey (DES, \cite{DES} provided for SGC).

As already mentioned, imaging systematics analysis is done on the North and the South separately. However, DESI DR1 clustering is measured separately on NGC and SGC and then combined appropriately (into GCcomb) \cite{DESI2024.II.KP3}.

\subsection{DESI emission line galaxies (ELGs)}
\label{subsec:elgs}
Emission Line Galaxies (ELGs) are star-forming and have become an important target for present and future spectroscopic surveys. 
Their importance lies in two facts \cite{elg_selection}. The first is that the star-formation rate density peaks at $z\sim$1-2, hence star-forming galaxies are common around these redshifts. Second, they show a distinct spectral feature, the [O II] doublet  3726,29$\,$\AA, that enables reliable and efficient spectroscopic redshift measurements. 
However, ELGs do not come without their challenges. In \cite{elg_selection} it has been shown that ELGs are sensitive to imaging systematics, due to their faint (low SNR) imaging. Thus, careful treatment of contamination  
should be performed on ELGs if any reliable cosmological information is to be extracted from the sample. 

The DESI DR1 ELG sample is defined in \cite{elg_selection}, and the selection cuts for the clustering catalog are defined in \cite{DESI2024.II.KP3} and summarized in Table \ref{tab:elg_sel}. In this paper, while we apply the same selection cuts as \cite{DESI2024.II.KP3} for ELGs $0.8<z<1.6$, we will also extend the sample selection to $0.6<z<0.8$ and test the imaging systematics. We apply the same `good' redshift criteria as \cite{DESI2024.II.KP3} for the entire extended redshift range of $0.6<z<1.6$. 

The DESI DR1 ELG sample spans over 5,924$\,\mathrm{deg}^2$ of sky, and is composed of 2,432,072 good redshifts within the $0.8<z<1.6$ range. This sample is split into two redshift ranges, $0.8<z<1.1$ and  $1.1<z<1.6$, which are also referred to as \elgo  and \elgt  respectively. 
As mentioned before, we extend the redshift range for ELGs to $0.6<z<1.6$, and focus on the sample within $0.6<z<0.8$, namely \elgz. These were not used in DESI 2024 cosmological analyses, as they showed greater systematic variation in number density as a function of redshift than the data at $z>0.8$, as we will detail in Section~\ref{subsec:dndz_obi}. The additional \elgz sample consists of 185,782 ELGs (within the same footprint as the $0.8<z<1.6$ sample), and we study the feasibility of recovering robust clustering from a sample with such strong imaging systematics.

\begin{figure}
    \centering
    \includegraphics[width=0.7\textwidth]{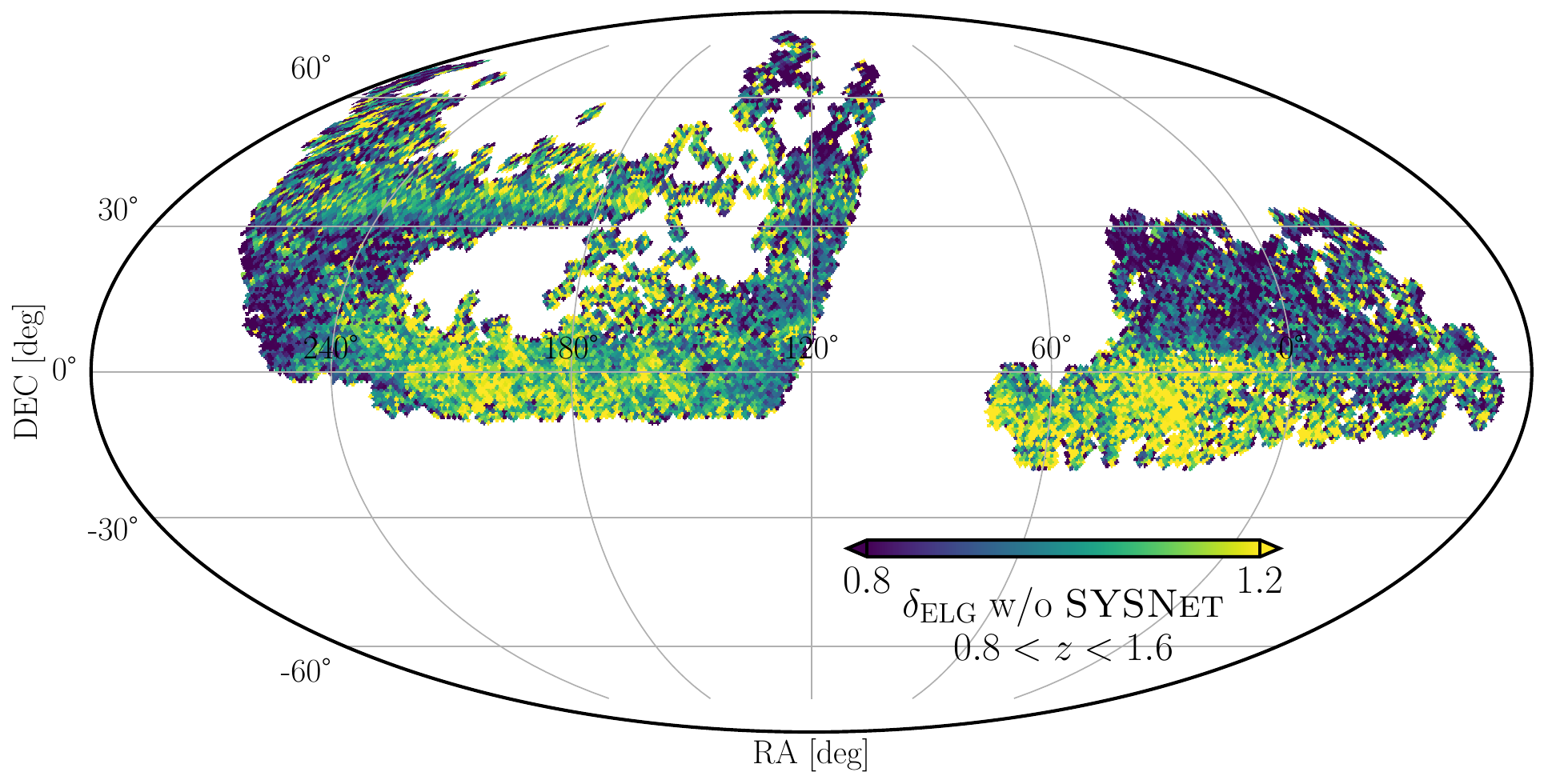}
    \includegraphics[width=0.7\textwidth]{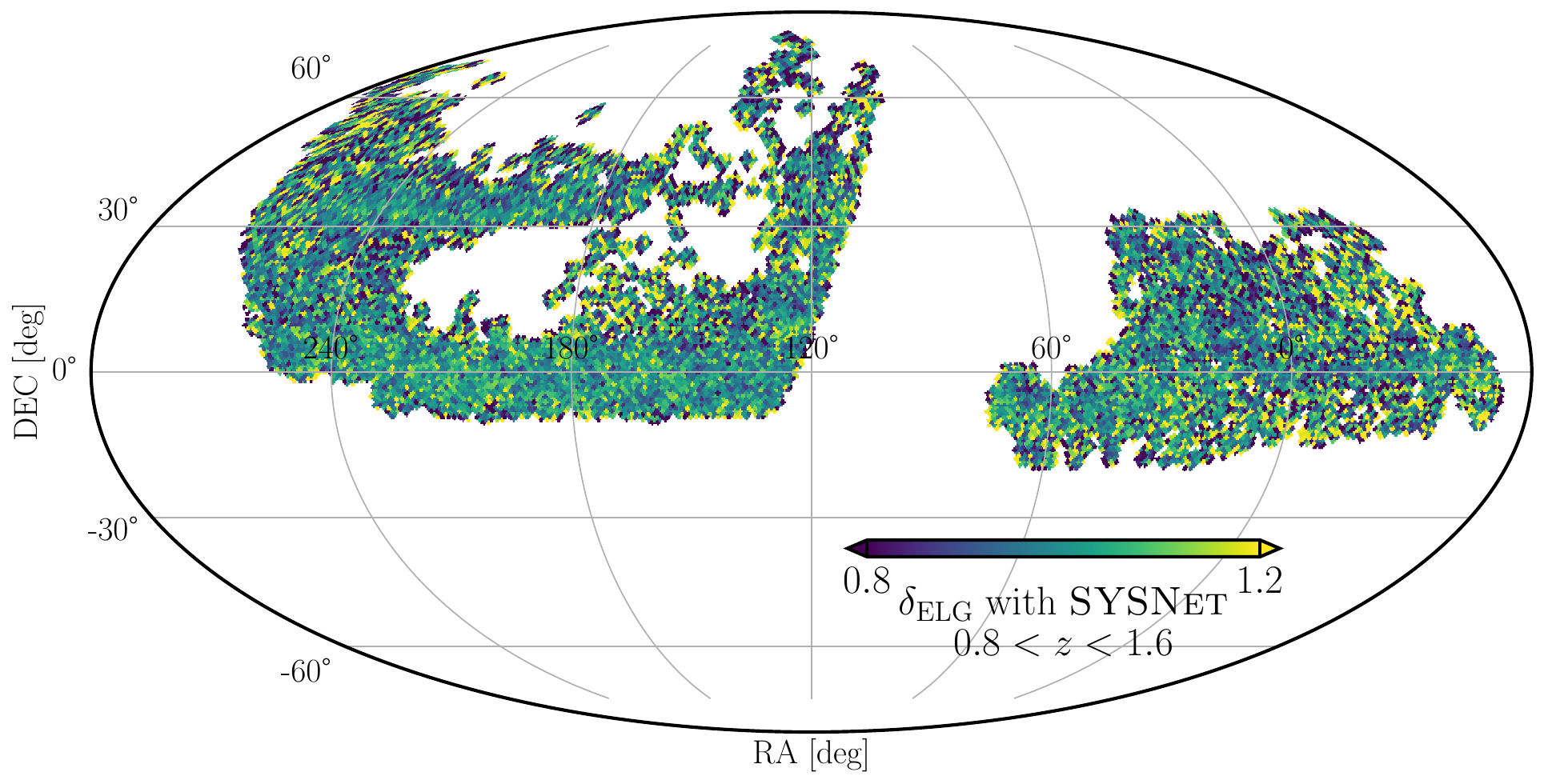}
    \caption{Projection of the normalized density of DESI DR1 ELGs, $\delta_\mathrm{ELG}$, within the $0.8<z<1.6$ redshift range. The top panel shows  $\delta_\mathrm{ELG}$ when no \sysnet weights are included, while the bottom panel shows  $\delta_\mathrm{ELG}$ after including \sysnet weights. When no systematic weights are applied we can see under- and over-dense regions in the sample caused by imaging systematics, e.g., the strong contrast in density observed 
    in the SGC region between over-dense for the DES region and under-dense for the non-DES region, which is likely due to the varying imaging depth for galaxies.}
    \label{fig:ELG_densmaps}
\end{figure}

\begin{figure}
    \centering
    \includegraphics[width=\textwidth]{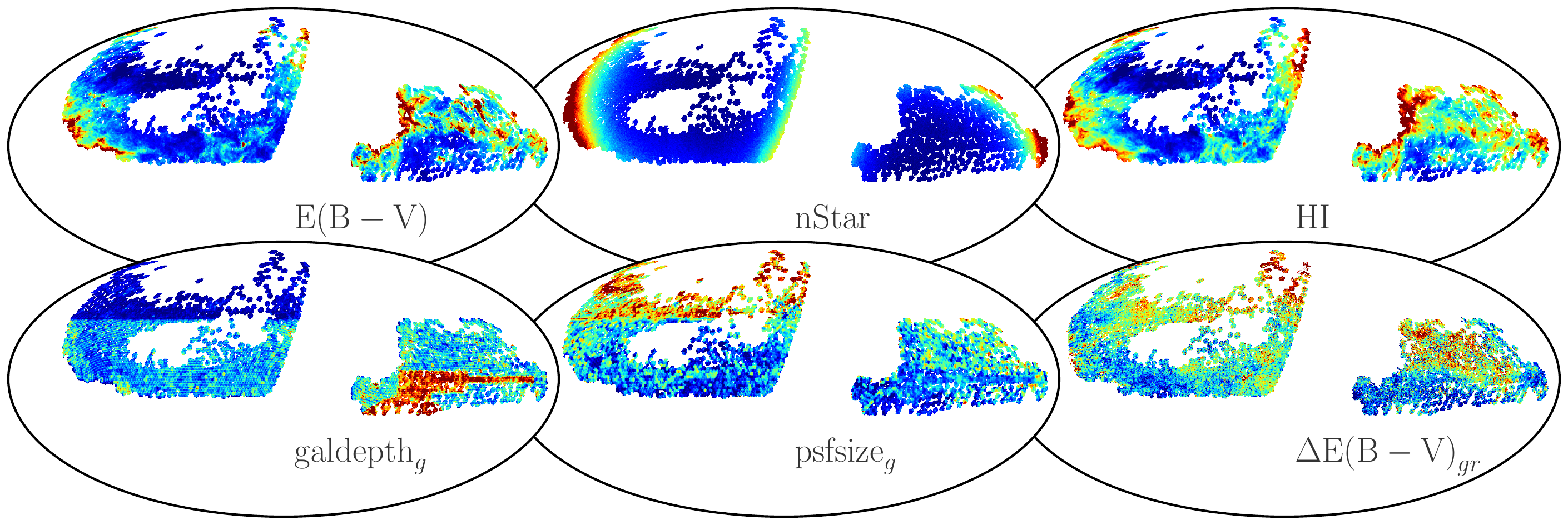}
    \caption{Mollweide projections of some of the systematic maps used throughout the paper in \textsc{HEALPix} with resolution of $NSIDE=256$. From left to right and top to bottom, we show galactic extinction from \cite{sfd98}, stellar density from \cite{GaiaDR2}, neutral hydrogen column density from \cite{HI.Bekhti}, imaging depth for galaxies and psfsize in the $g$-band, and $\Delta\mathrm{E(B-V)}_{gr}$ from \cite{KP3s14-Zhou}.}
    \label{fig:sysmaps}
\end{figure}

In Fig. \ref{fig:ELG_densmaps} we show sky maps for the observed DESI DR1 ELG density normalized by its mean ($\delta_\mathrm{ELG}$), before (top panel) and after (bottom panel) mitigating imaging systematics. 
Imaging systematics for DESI ELGs are mitigated using the \textsc{SYSNet}\footnote{https://github.com/mehdirezaie/sysnetdev} pipeline, and details are presented in Section~\ref{sec:sysnet}. 
In the un-mitigated sample we can see that there are density fluctuations across the sky. For example, if we direct our focus towards the SGC region, then we can observe over- and under-densities in the DES and non-DES region respectively that seem to correlate with the depths in these regions (see Fig.~\ref{fig:sysmaps}). Qualitatively, the application of systematic weights corrects for the fluctuations and the sample appears more uniform. The remaining density fluctuations do not appear to be an imprint from any known systematic.

\begin{table*}
    \centering
    \begin{tabular}{cccc}
    \hline
    \hline
         Sample&  Cuts& Comments\\
    \hline
         &    $(g>20)$ and $(g_\mathrm{fib}<24.1)$& Magnitude cut\\
         ELG\_LOP&  $0.15<r-z$& $r-z$ cut \\
         &    $g-r < 0.5 \times (r-z) + 0.1$& Star/low-z cut\\
         &    $g-r < -1.2 \times (r-z) + 1.3$& Redshift/$[\mathrm{O}\, \mathrm{II}]$ cut\\
    \hline
    \end{tabular}
    \caption{DESI ELG target selection cuts from \cite{elg_selection}, magnitudes are corrected for Galactic extinction. These cuts as described in \cite{elg_selection} are what is used for DR1 `low' priority ELG (\texttt{ELG\_LOP}) selection, which is the ELG sample used for DR1 cosmological analysis. The $g>20$ cut discards bright objects that are unlikely to be $z>0.6$; the $g_\mathrm{fib}<24.1$ is a faint cut such that the desired \texttt{ELG\_LOP} density is reached by design. The $g-r$ vs. $r-z$ cuts are at first redshift cuts that favor $1.1<z<1.6$, and secondly optimize detection of [O II] emitters.}
    \label{tab:elg_sel}
\end{table*}

\subsection{Other DESI DR1 tracers}
In this section we briefly describe the other DESI DR1 tracers used in the DR1 analysis, which in this paper we use to compare clustering statistics with mitigated and un-mitigated systematics, and also test the BAO feature against systematic contamination.
\begin{itemize}
    \item Luminous red galaxies (LRG): These are elliptical galaxies with old stellar populations, hence they are red. The DESI DR1 sample is composed of 2,138,627 good redshifts objects within the $0.4<z<1.1$ redshift range, covering 5,740$\,\mathrm{deg}^2$. For cosmological analysis, this sample is subdivided into \lrgo, \lrgt and \lrgth. The fiducial mitigation scheme for this sample is based on linear regression and is described in \cite{DESI2024.II.KP3}.

    \item Quasars (QSO): These consist of 856,831 good redshifts objects within the $0.8<z<2.1$ redshift range, covering 7,249$\,\mathrm{deg}^2$. The imaging systematics for this sample were carefully mitigated with a random forest based approach called \regressis, developed and described in \cite{Chaussidon21QSOsys}.
    
    \item Bright Galaxies (BGS): These include 300,043 good redshifts objects within the $0.1<z<0.4$ redshift range, covering 7,473$\,\mathrm{deg}^2$. The treatment of imaging systematics for this sample is also linear regression based and is described in \cite{DESI2024.II.KP3}.
\end{itemize}

\subsection{Imaging attributes}
During imaging observations various conditions introduce non-cosmological fluctuations in the ELG target density across the footprint. These imaging conditions have unknown correlations with the target density and propagate to the spectroscopic sample. These imaging attributes are due to foreground contamination or of instrumental nature and can spatially and temporally impact image quality. In this work imaging attributes are contained in \texttt{HEALPix} maps of $\mathrm{NSIDE}=256$ resolution, which we also refer to as imaging maps or imaging templates hereafter. Here we briefly describe the imaging attributes used throughout the paper (a more detailed description on the construction of these maps is found in the Appendix of \cite{DESI2024.II.KP3}):

\begin{enumerate}
    \item \textbf{Galactic extinction}: labeled as $E(B-V)$ (also referred to as SFD98), with units of magnitude (mag). This map uses the thermal emission of dust measured at wavelengths of $100\,\mathrm{\mu m}$ and longer \cite{sfd98} as an estimator of galactic reddening. 
    
    \item \textbf{Stellar density}: labeled as $\mathrm{nStar}$, with units stellar density per square degree ($\mathrm{deg}^{-2}$). This map is made from Gaia DR2 \cite{GaiaDR2} stars within $12\leq G<17$.
    
    \item \textbf{Point spread function size}: labeled as $\mathrm{psfsize}_{b}$ ($b = grz$-bands), 
    full-width half maximum (FWHM) of the point spread function (PSF) in arcseconds ($\mathrm{arcsec}$) \cite{jarvis2021dark,bertin2011automated}.
    
    \item \textbf{Image depth for galaxies}: labeled as $\mathrm{galdepth}_{b}$ ($b=grz$-bands), $5\sigma$ galaxy detection depth in units of magnitude (mag). All galactic depths in this paper are corrected for dust extinction. For conciseness, we refer to this property as `galaxy depth' throughout the paper.
    
    \item \textbf{Hydrogen column density}: labeled as \texttt{HI}, with units of $\mathrm{cm}^{-2}$. This map combines data from the Effelsberg-Bonn H\,I Survey and the third revision of the Galactic All-Sky Survey to generate an all-sky map of neutral hydrogen column density \cite{HI.Bekhti}.
    
    \item \textbf{DESI galactic extinction}: labeled as $E(B-V)_c$ where $c$ corresponds to colors ($g-r$ and $r-z$), with units of mag. These are the Galactic extinction maps derived in \cite{KP3s14-Zhou} using DESI stars. 
    
    \item \textbf{Difference between DESI $E(B-V)_c$ and $E(B-V)$}: labeled as $\Delta E(B-V)_c$, where $c$ corresponds to colors ($g-r$ and $r-z$), with units of mag. This map contains the difference between DESI $E(B-V)_c$ and $E(B-V)$.
\end{enumerate}

In Fig.~\ref{fig:sysmaps} we show the projection of a few of the systematics maps used for our imaging systematics analysis. By comparison, we can qualitatively see imprints of these systematics in the density variations observed in the un-mitigated DESI DR1 ELG sample (see Fig.~\ref{fig:ELG_densmaps}). Our goal is to reliably remove these imprints from the observed ELG sample, and assess the remaining systematics.

\section{Analysis methods}
\label{sec:analysis_methods}
\subsection{Mean galaxy density}
We measure the projected density of ELGs as a function of imaging properties and use it as a primary diagnostic for the presence of systematic variation. Any trend found in such measurements suggests systematic contamination, as the true galaxy density should not depend on, e.g., the observing conditions. 
We construct the mean galaxy density, $\delta$, as function of an imaging property attribute, $a_j$, similar to \cite{Rezaie_2021},
\begin{equation}\label{eq:overdensity}
    \delta(a_j) = \alpha \frac{\sum_i n_{g,i}(a_j)}{\sum_i n_{r,i}(a_j)} - 1.
\end{equation}
Where $n_{g,i}$ and $n_{r,i}$ are the weighted number of galaxies and randoms in pixel $i$; $\alpha$ is the factor to normalize the number of randoms to that of galaxies, which we calculate as the total number of weighted galaxies over the total number of weighted randoms; and the summation of $\sum_i$ is computed over pixels with $a_{j,i}\in [a_j,a_j+\Delta a_j]$, where $\Delta a_j$ are linearly spaced bins in the range between the $0.1\mathrm{th}$ and $99.9\mathrm{th}$ percentile of $a_j$. We assign errors on the mean galaxy density as follows, which use a form assuming Poisson error that incorporates weights based on the mean density to approximately account for cosmic variance as done in \cite{DESI2024.II.KP3, RossDR12},
\begin{equation}\label{eq:sigma_dens}
    \sigma(a_j) = \alpha \frac{\sqrt{\sum_i n_{g,i}}}{\sum_i n_{r,i}}.
\end{equation}
For the DESI DR1 data, in addition to the completeness and systematic weights described in Eq. \ref{eq:weights}, we include weights that account for the variation of the completeness and FKP weights with the number of tiles $n_\mathrm{tile}$, namely $w_\mathrm{FKP,2D}$ (details in \cite{DESI2024.II.KP3}).

The mean galaxy density measurements as determined above do not include any direct positional information, as these are averaged over the entire footprint. 
In our study, we opted to match the pixel resolution of the DESI imaging systematic maps as \cite{DESI2024.II.KP3}, i.e. $\mathrm{NSIDE}=256$ which corresponds to pixels with a side length of 0.2290 degrees. 

\subsection{Clustering measurements}
The DESI collaboration extracts cosmological information from its LSS catalogs by measuring the spatial distribution of galaxies through anisotropic 2-pt statistics. Clustering measurements are done in configuration and Fourier-space. In configuration space we compute the 2-pt correlation function using \textsc{pycorr}\footnote{\url{https://github.com/cosmodesi/pycorr}}, while for its Fourier counterpart the power spectrum we use \textsc{pypower}\footnote{\url{https://github.com/cosmodesi/pypower}}. For these computations we follow \cite{DESI2024.II.KP3}, where they describe in detail the codes and methods used to obtain clustering measurements across all DESI DR1 companion papers.

In this work we produce variations of the 2-pt measurements to assess the impact of different choices of weighting schemes. Thanks to how the clustering catalogs were created \cite{KP3s15-Ross}, we can easily ignore or add the effect of weights by dividing or multiplying the \texttt{WEIGHT} column by the column of interest. If the weight column we desire to add or remove is not contained in the clustering catalog or the `full' catalogs, we first add the column to the `full' catalog, then the column is added to the clustering catalog by matching the \texttt{TARGETID} of the clustering catalog to the `full' catalog. Below we list the variations of the 2-pt measurements for the ELGs shown in this paper (all variations have their respective completeness and FKP weights applied, as in \cite{DESI2024.II.KP3}):
\begin{itemize}
    \item $\mathrm{raw}$: Measurements obtained when ignoring systematic weights. 
    
    \item $\mathrm{SN}$: Measurements obtained when weighting by angular systematic weights obtained from \textsc{SYSNet}, as described in section \ref{sec:sysnet}. 
    
    \item $\mathrm{SN}+N(z)$: Measurements obtained when weighting by angular systematic weights obtained from \textsc{SYSNet}, and adding radial systematic weights obtained in section \ref{subsec:dndz_obi}. 
    
    \item $\mathrm{SN}+N(z)_\mathrm{\Delta EBV}$: Measurements obtained when weighting by angular systematic weights obtained from \textsc{SYSNet}, and adding radial systematic weights obtained in section \ref{subsec:dndz_ebv}. 
    
    \item $\mathrm{obiNN}$: Measurements obtained when weighting by angular systematic weights obtained from the combined \textsc{Obiwan}+\textsc{SYSNet} approach described in section \ref{sec:obinn}. 
\end{itemize}
When presenting 2-pt measurements for the other tracers, we only consider the `raw' and the default imaging weight as defined in \cite{DESI2024.II.KP3}.

\section{Regressing against imaging systematics with SYSNet}
\label{sec:sysnet}
Early in the analysis of the DESI DR1 ELG sample, it became apparent that the impact of imaging systematics was such that any linear regression technique would not suffice. We therefore
adopted \sysnet as the regression method to apply to the DR1 ELG sample.
The \sysnet pipeline is a fully connected Feed Forward Neural Network developed and validated by \cite{Rezaie_2020}, which models the relationship between imaging templates (imaging maps) and the observed galaxy density field. The pipeline output is then used to generate a selection mask that up-weights (we multiply by the normalized inverse predicted galaxy counts)  the observed density field, reducing the spurious trends between the imaging systematics and the observed target density. \sysnet has been used to prepare eBOSS DR16 Quasars \cite{Rezaie_2021} and DESI LRGs \cite{rezaie2023local} for primordial non-Gaussianity analysis, and to prepare DR9-LS ELGs \cite{karim2024measuringsigma8usingdesi} for $\sigma_8$ measurement in cross-correlation with CMB lensing.

\subsection{Description of SYSNet}
\sysnet is a regression based approach that uses a neural network to model the dependence of the observed galaxy density field on different imaging features. It is important to note that \sysnet does not make any assumptions about linearity, nor does it know about angular position (RA and DEC are not used as input) when training the model. Since the regression is designed to work at the pixel level, the observed galaxy counts (\emph{label}) and the imaging systematics (\emph{features}) have to be \textsc{HEALPix} \cite{healpix} maps of matching resolution. 

In \cite{Rezaie_2021} it was found that a Poisson statistics based cost function, compared to a Mean Squared Error (MSE) cost function penalized by a L2 regularization \cite{ridge_regression} term used in \cite{Rezaie_2020}, improves the predictive performance of \sysnet since it better accounts for the counting of sparse samples of galaxies and asymptotically approaches the MSE case as the density increases. 
Hence, we also use the negative Poisson log-likelihood as the cost function for the regression. This assumes that the galaxy counts in a given pixel are independent from the galaxy counts in other pixels\footnote{This is not strictly true since there is large-scale structure.} and follows a Poisson distribution. In a per pixel formalism, where $i$ is used to represent the $i$-th pixel, we refer to the observed galaxy counts per pixel as $y_i$ and the vector of imaging systematics as $\mathbf{a}_i$. We then write the likelihood $\mathcal{L}$ as the joint probability for $N$ pixels
\begin{equation}
    \mathcal{L} = \prod_{i=1}^N f(y_i \vert \theta,\mathbf{a}_i),
\end{equation}
where $\theta$ is the parameter vector. If the galaxy counts follow a Poisson distribution,
\begin{equation}
    f(y_i \vert \theta,\mathbf{a}_i) = \frac{\lambda_i^{y_i}\exp^{-\lambda_i}}{y_i !},
\end{equation}
where $\lambda_i \equiv \lambda(\theta,\mathbf{a}_i)$ is the expected number of galaxies per pixel as a function of the imaging systematics. Then the goal is to find the bestfit parameters that maximize $\mathcal{L}$, or minimize the negative logarithm of $-\log\mathcal{L}$, which is the cost function we use,
\begin{equation}
	-\log(\mathcal{L}) = \sum^N_{i=1}\left[\lambda_i - y_i \log(\lambda_i) \right].
\end{equation}

\sysnet has the option for $k$-fold cross-validation. We use $k=5$ cross-validation such that the data is randomly split into $k$ equally sized folds or groups. The distribution of the data is then: three folds ($60\%$) used for training the model, one fold ($20\%$) for validation and one fold ($20\%$) for the test set. Furthermore, 5 partitions of the data are created by the random permutation of the folds, such that one can recover the full data vector if we merge the test fold of each partition. In other words, each fold is used once as a test set in each partition, which allows the NN model to be tested on the entire data set. Fig 4 in \cite{Rezaie_2020} shows a visualization of how the data is partitioned under this five-fold cross-validation. 
The training fold is then split into batches of size $N_\mathrm{batch}$; splitting into batches reduces memory usage, and helps the training stability and convergence rates \cite{batch_normalization}.
$N_\mathrm{batch}$ batches are processed once each epoch, updating the NN parameters each epoch. Also, in each epoch the NN model is applied to the validation set to assess the predictive accuracy of the model. The test set is then used for the application of the NN model with lowest validation error after being trained for $N_\mathrm{epochs}$ number of epochs. 

In \cite{Rezaie_2021}, to reduce the noise on the predicted galaxy counts from a single NN prediction model, they generate an ensemble of NN models by randomly initializing the NN parameters. In this work we refer to the NN models with random initializations as chains. The amount of chains that \sysnet creates we call $N_\mathrm{chains}$ and can be specified by the user. As an example, $N_\mathrm{chains}=5$ with five-fold cross validation would produce an ensemble of 25 NN models. Then the final NN prediction or selection function is obtained by taking the mean of the predicted galaxy counts across the ensemble of NN models. We then represent the mean of this ensemble predicted galaxy counts per pixel as $\bar{\eta}(\theta,\mathbf{a}_i)$, where the bar is for the ensemble average over NN models. $\bar{\eta}(\theta,\mathbf{a}_i)$ is also called the selection function, from it we generate photometric weights at the pixel level by
\begin{equation}\label{eq:wsys}
    w_\mathrm{SN} = \frac{1}{\bar{\eta}(\theta,\mathbf{a}_i)},
\end{equation}
which can be used to mitigate the observed galaxy density field by multiplying the observed galaxy counts by $w_\mathrm{SN}$. Please note that these weights are normalized by their mean value to keep the total number of weighted galaxies the same as unweighted galaxy counts; to avoid extreme predictions, we then clip the weights to be between [0.5, 2.0]\footnote{The fraction of weights clipped in this way is small for all samples. Specifically, for the \elgo North and South 0.013\% and 0.004\% respectively; while for the \elgt North and South 0.018\% and 0.003\% respectively.}.

\subsection{SYSNet for DESI ELGs}
\label{subsec:sysnet_elgs}
Here we describe how the fiducial systematic weights for DESI DR1 ELGs are obtained and specify how they are added to the LSS catalog. The DESI DR1 ELG sample spans over $0.8<z<1.6$, and for NN training it has been divided into two redshift bins, $0.8<z<1.1$ (hereafter, \texttt{ELG1}) and $1.1<z<1.6$ (hereafter, \texttt{ELG2}). Furthermore, each of the samples is further divided into subsamples based on the two distinct photometric regions, North and South.  
This results in four data sets: \elgo North, \elgo South, \elgt North, and \elgt South.
Details of hyper-parameters used for the \sysnet pipeline are presented in \cref{sec:SN_settings}.

\sysnet takes input data prepared in a \textsc{FITS}\footnote{https://fits.gsfc.nasa.gov/fits standard.html} file containing the following columns: observed galaxy counts per pixel, \textsc{HEALPix} pixel number, pixel completeness, and imaging maps for which we desire to model the correlations with the data. 
We use 10 imaging maps for training, and they are as follows: nStar, \texttt{HI}, $\mathrm{psfsize}_{\{g,r,z\}}$, $\mathrm{galdepth}_{\{g,r,z\}}$, $\Delta E(B-V)_{\{gr,rz\}}$.
The HEALPix maps used are in ring ordering with resolution $\mathrm{NSIDE}=256$ which corresponds to pixels with a side length of 0.2290 degrees. The galaxy counts are weighted by
\begin{equation}\label{eq:weights}
	w = w_\mathrm{comp} \times w_\mathrm{zfail},
\end{equation}
where $w_\mathrm{comp}$ is the inverse of the completeness statistics used to account for completeness variations, and $w_\mathrm{zfail}$ accounts for changes in the relative success rates \cite{DESI2024.II.KP3}. To be more specific $w_\mathrm{comp} = 1/(f_\mathrm{TLID}f_\mathrm{tile})$, where $f_\mathrm{TLID}$ is a per tile location completeness, and $f_\mathrm{tile}$ is a per tile group completeness (a detailed description of these completeness weights is in \cite{DESI2024.II.KP3}). These choices were made so that no weights needed to be applied to the randoms, which have a uniform density in DESI LSS catalogs. This allows
the pixel completeness to be computed by finding the fraction per pixel of ELG randoms to the full sky randoms.
Since geometry and veto masks are equally applied to data and randoms, the pixel completeness we compute not only accounts for imaging masks but also survey geometry, hardware vetoes, and imaging properties masks (all relevant masks and vetoes are described in Section~4 of \cite{DESI2024.II.KP3}).
These values are important for weighting the NN output according to the fraction of galaxies and randoms in each pixel.

\begin{figure}
    \centering
    \includegraphics[width=\textwidth]{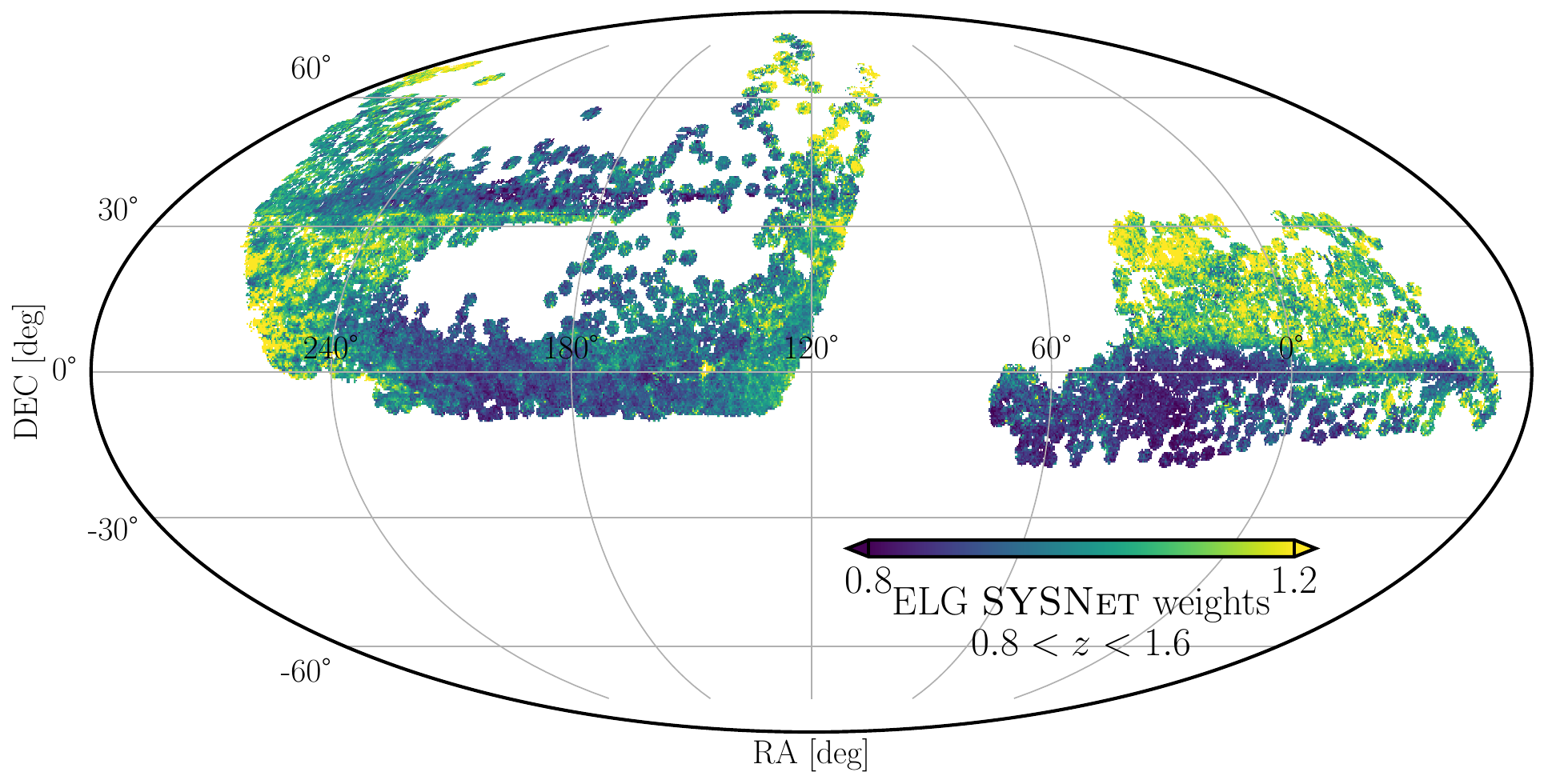}
    \caption{Map of the \sysnet weights derived for DESI DR1 ELGs withing the $0.8<z<1.6$ redshift range. This map was generated by projecting the $w_\mathrm{SN}$ at the catalog level to a \textsc{HEALPix} map of $N_\mathrm{side}=256$ resolution.}
    \label{fig:sn_densmaps}
\end{figure}

After running the \sysnet pipeline, as described above, we generate a set of systematic weights $w_\mathrm{SN}$ by taking the mean of the predicted galaxy counts over the ensemble of 25 NN models. We then project these systematic weights per pixel to a \textsc{HEALPix} map with a resolution that matches the input resolution, $\mathrm{NSIDE}=256$. Then this map of systematic weights is used to assign a systematic weight to each object in the data catalog.
Fig.~\ref{fig:sn_densmaps} shows the systematic weight sky-map obtained for DESI DR1 ELGs within $0.8<z<1.6$. Qualitatively, compared to the sky-map of the observed ELG density before any mitigation shown in Fig.~\ref{fig:ELG_densmaps}, the weights obtained using \sysnet captures the overall ELG density fluctuations across sky, up-weighting and down-weighting where the ELG density is lower and higher than average  respectively. The angular effect of applying the systematic weights $w_\mathrm{SN}$ is already shown in  Fig.~\ref{fig:ELG_densmaps}.

\section{Forward modeling systematic variation using Obiwan}
\label{sec:obiwan}
\obiwan is an image simulation tool, developed by \cite{obiwan}, that injects synthetic galaxies measured from deep photometry into survey images. 
The modified images are later processed in the same way as real images to ensure similar observing conditions for synthetic galaxies. Therefore, \obiwan preserves most of the code structure from \textsc{Legacypipe}\footnote{Pipeline designed for the detection and characterization of sources in the Legacy Survey (\hyperlink{https://github.com/legacysurvey/legacypipe}{https://github.com/legacysurvey/legacypipe}).}, the image processing pipeline used to generate catalog level data from real survey images.
Since the number of injected sources is known, it can extract how many of the fake sources were lost and recovered. Checking the recovery fraction is just one aspect. \obiwan also simulates how the observed flux and shape of injected galaxies are perturbed when extracted from real images.
Assuming that the fake sources have been properly injected into the images and that the images themselves contain all of the sources of systematic variation, the process can be used to generate a simulated catalog that suffers from the same systematic variation in observed galaxy counts as the real data \cite{obiwan}. 

\subsection{Description of Obiwan} 
The \obiwan pipeline, like \textsc{Legacypipe}, processes images at the brick level. For the Legacy Survey, images are gridded with a pixel scale of $0.262"$ and $3600\times3600$ pixels form a brick, hence a `brick' is a $0.25\deg \times 0.25\deg$ piece of sky defined in RA and Dec coordinates. 
\obiwan injects sources within a brick by placing the synthetic galaxies in a hexagonal pattern with each galaxy separated by $37.728"$ (144 pixels), this hex-grid method is developed and used in \cite{KP3s13-Kong}. In \cref{sec:obiwan_appendix} we present any relevant computational details.

\obiwan was run for the complete DESI Y1 sample, this was done separately for the North and the South. We first ran for the bricks contained in observations from the first two months (M2; \cite{BAO.EDR.Moon.2023}) of the DESI main (North: 13,874, South: 29,272) sample, which were completed using the Cori machine at NERSC; the remaining bricks needed to complete the area covered in DR1 (North: 19,599, South: 96,022) footprint were later completed using the Perlmutter machine at NERSC. Due to errors or long computation time, 370 and 1,128 bricks were unfinished for the North and South respectively\footnote{The exclusion of these bricks does not impact our analysis.}.

In this work, we produce synthetic DESI ELGs with \textsc{Obiwan}\footnote{\hyperlink{https://github.com/adematti/legacysim.git}{https://github.com/adematti/legacysim.git}}. The target selection of DESI ELGs involves 3 optical bands: $g$, $r$ and $z$ band. Therefore, the simulation of \obiwan ELGs is also performed on the same 3 optical bands. Unlike the simulation of \obiwan LRGs \cite{KP3s13-Kong}, which utilizes a WISE band with large PSF, \obiwan ELGs do not require simulation in WISE. WISE has a higher level of blending which limits the number density of injected galaxies in each iteration. Avoiding the WISE band ensures that the injection of \obiwan ELGs can reach a much higher number density than \obiwan LRGs. 

\obiwan models galaxies as several types: Point source (PSF), Round exponential galaxies (REX), Exponential galaxies (EXP), de Vaucouleurs galaxies (DEV) and S\`ersic galaxies (SER). The galaxies are convolved with the point spread function measured at the injected location for each individual location. We use \textsc{Legacypipe} to produce these convolved stamps. This is equivalent to using \textsc{Galsim} \cite{rowe2015galsimmodulargalaxyimage}, which is a more commonly used pipeline for producing synthetic galaxies. Compared with \textsc{Galsim}, \textsc{Legacypipe} utilizes a more efficient method for galaxy-PSF convolution \cite{lang2020hybridfourierrealgaussianmixture}.

We select ELGs from the dr9-COSMOS deep sample. The galaxies in this sample are measured with $\sim$ 10 times more images for each band, compared to the nominal data used for DESI ELG targeting, so their flux error is much smaller. We treat it as the `truth' sample. We select an ELG-like sample from the `truth sample' with an extended color selection. 
The color selection is enlarged by $1.2\,\mathrm{mag.}$ in the $g$-band, $0.8\,\mathrm{mag.}$ in the $g-r$ color, and $0.5\,\mathrm{mag.}$ in the $r-z$ color. 
We inject these ELG-like galaxies into the DR9 images and run \textsc{Obiwan} to obtain the output sample. Finally, we apply the DESI ELG color selection function \cite{elg_selection} on the output galaxies to obtain the \obiwan ELG sample. 

\subsection{Trends predicted with Obiwan ELGs}
We use the \obiwan pipeline to simulate the entirety of the DR1 ELG sample, and assess how well it is able to simulate systematic trends observed in DESI ELGs. 

In the \obiwan simulations, sources were injected assuming SFD98 reddening. However, improved dust maps are available to us from \cite{KP3s14-Zhou}. They showed that a target sample selected with photometry dereddened by DESI \ebv maps is significantly more uniform than when the SFD98 \ebv map is used for extinction correction. Hence, we have to implement this extinction correction at a post-processing stage. Running multiple \obiwan simulations with different assumed \ebv would be prohibitive based on the computational expense (see \cref{sec:obiwan_appendix} for runtime details). Instead, we emulate the effect of a change in the assumed \ebv by altering the photometry of the output catalog. ELGs are selected following Table \ref{tab:elg_sel}. The photometry is altered in the following way, replacing the extinction of SFD98 with the extinction model from \cite{KP3s14-Zhou}, which yields
\begin{equation}
    \begin{aligned}
        r-z &= r-z + (c_r-c_z)\times \Delta{E(B-V)_{color}} \\
        g-r &= g-r + (c_g-c_r)\times \Delta{E(B-V)_{color}} \\
        g & = g + c_g \times \Delta{E(B-V)_{color}} \\ 
        g_\mathrm{fib} & = g + c_g \times \Delta{E(B-V)_{color}}
    \end{aligned}
\end{equation}
where $\Delta{E(B-V)_\mathrm{color}} = E(B-V)_\mathrm{color} - E(B-V)_\mathrm{SFD}$, with $E(B-V)_\mathrm{SFD}$ from SFD98 and $E(B-V)_\mathrm{color}$ from \cite{KP3s14-Zhou}, and $c_b$'s are the extinction coefficients corresponding to each band ($c_g=3.214$, $c_r=2.165$, $c_z=1.211$ \cite{schlafly2011measuring})\footnote{Extinction values from \hyperlink{https://www.legacysurvey.org/dr10/catalogs/}{https://www.legacysurvey.org/dr10/catalogs/}}. There are two versions of improved dust maps, $rz$ and $gr$ color versions. For this work we will primarily compare to the results that use $E(B-V)_\mathrm{rz}$ (both maps are used in regressions).

\begin{figure}
    \centering  
    \includegraphics[width=0.9\textwidth]{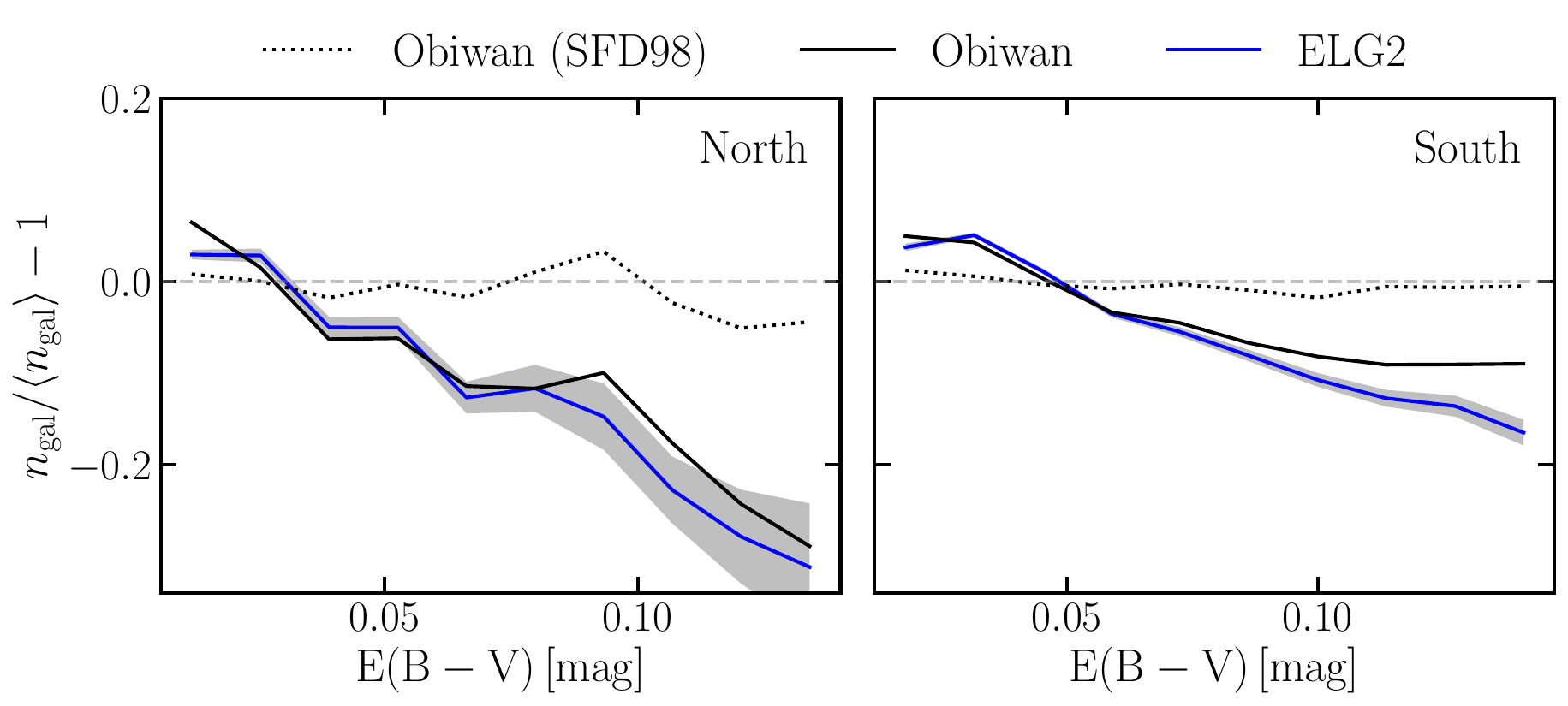}
    \caption{Relative mean number density of DESI DR1 ELGs and \obiwan ELGs as a function of the SFD98 \ebv feature. Left panel corresponds to the North ELG sample, right panel shows the South sample. Solid blue line corresponds to DESI ELGs, the dotted black line corresponds to the \obiwan sample, and the solid black line shows the \obiwan sample after implementing the DESI dust map. \obiwan using SFD98 values for reddening does not provide accurate predictions of ELG trends with important features.
    }
    \label{fig:dens_obi_ebv}
\end{figure}

In Fig \ref{fig:dens_obi_ebv}, we show the relative ELG density as a function of the \ebv SFD98 map \cite{sfd98} for the the North (left-panel) and the South (right-panel). The solid blue line represents the \elgt sample and has a grey shadowed region that shows its $1\sigma$ error, see Eq. \ref{eq:sigma_dens}. 
The black solid lines represent the \obiwan sample that underwent target selection with the DESI \ebv map. 
The black dotted curve represents the \obiwan sample that underwent target selection with the \ebv SFD98 extinction values.  
The uncertainty on \obiwan trends is based purely on assuming Poisson variations in the counts. 
Ideally, \obiwan would replicate the fluctuations seen in the observed ELG sample. This is mostly true for the DESI\ \ebv selected sample (black solid lines). This sample shows a strong improvement, as expected, over the SFD98 selected sample. The improvement is shown by the overall consistency, within $1\sigma$ error most of the time, between the \obiwan target density (solid black lines) and the observed DESI ELGs (blue lines). However, there is still some unknown systematic at $E(B-V)=0$ for both samples and in the South sample for $E(B-V)>0.8\,\mathrm{mag}$.

\begin{figure}
    \centering
    \includegraphics[width=\textwidth]{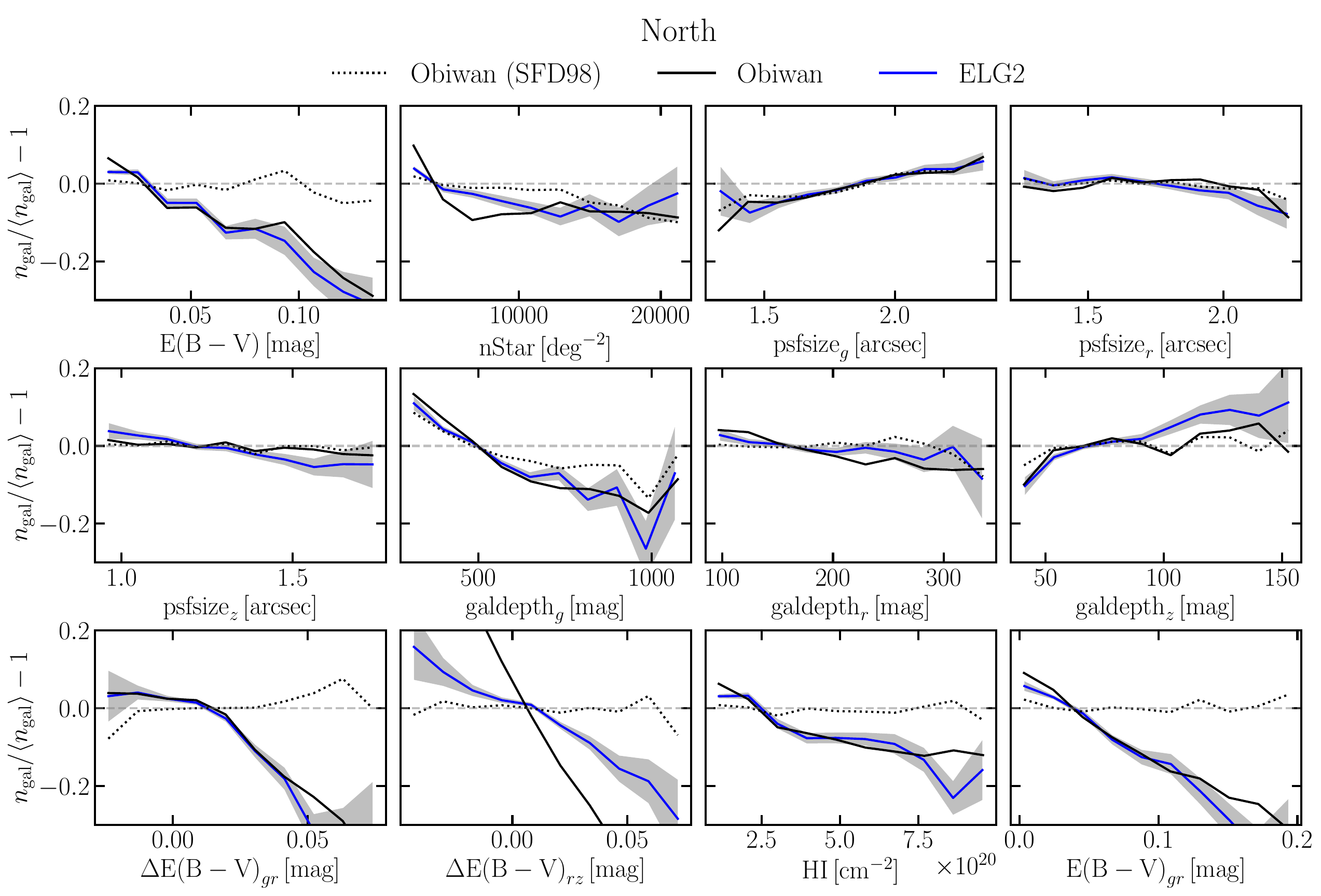}
    \caption{Galaxy mean density as a function of systematics for \elgt Nortrh. Solid blue lines correspond to DESI ELGs, while the black lines show \obiwan ELGs before (dotted lines) and after (solid lines) using an improved dust map.}
    \label{fig:obi-dens-N}
\end{figure}

\begin{figure}
    \centering
    \includegraphics[width=\textwidth]{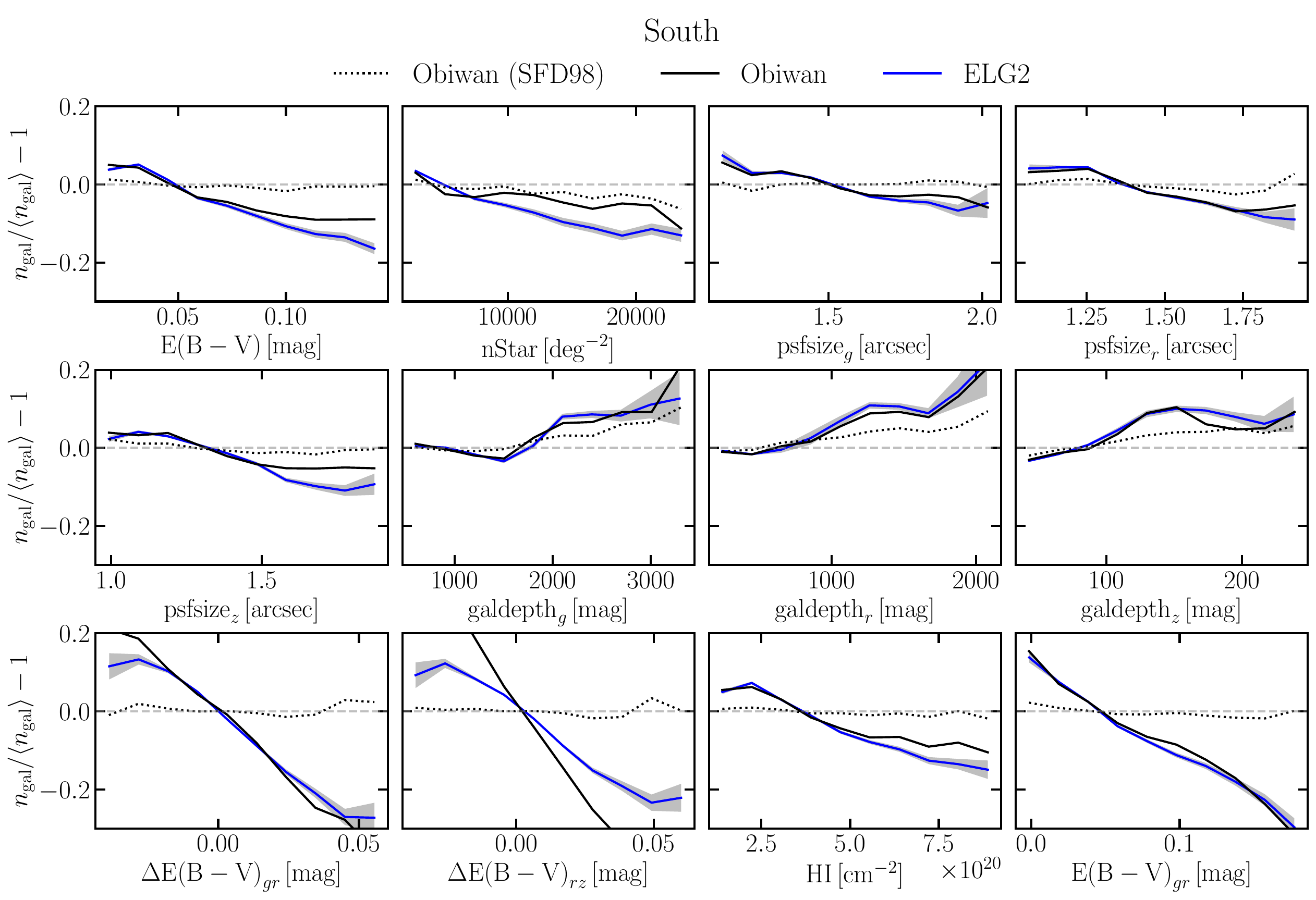}
    \caption{Same as Fig. \ref{fig:obi-dens-N}, but for \elgt South.}
    \label{fig:obi-dens-S}
\end{figure}

In Figs.~\ref{fig:obi-dens-N} and \ref{fig:obi-dens-S} we show the results for all of the imaging maps considered in our analysis.
Overall, in both regions, we find that the DESI \ebv corrected \obiwan results show trends consistent with the observed ELGs for the majority of the imaging attributes presented.
However, it is important to note that the trends for the South are indeed improved but there still seems to be some unknown systematic that is not present in the North. These discrepancies can be because of unknown factors that are not simulated by \textsc{Obiwan}. Also, \obiwan galaxies are not exactly the same as real galaxies.
Directing our attention to the $\Delta E(B-V)_{rz}$ subplot, we see that the trends seem to be over-predicted, i.e. the trend between \obiwan and the feature is greater than the one observed in the ELG sample and the feature, \obiwan shows a steeper negative slope in its trend. The over-prediction is likely due to noise in the \ebv maps, i.e., the values in the map look like $E(B-V)_\mathrm{true} + \epsilon$, where $\epsilon$ represents some scatter. The scatter places more results at extreme $\Delta E(B-V)$ and then \obiwan predicts a more extreme trend. Given the $\Delta E(B-V)$ is the strongest overall trend we expect that if we had applied a `true' \ebv map, the trends against all maps would also improve. We leave this for future investigation but conclude that it is clear that the change in extinction values predicted in \cite{KP3s14-Zhou} is the dominant source of systematic variation in the ELG target density. 

We also observe a mismatch in the stellar density trend. One likely reason for the remaining mismatch of \obiwan ELGs in the stellar density trend is the zero-point calibration. \cite{Zhou_2023} discusses the zero-point calibration error in Legacy Survey.
ELGs are impacted by the zero-point calibration error.
The zero-point error is estimated from Gaia star catalog, which also has zero-point calibration error. In Fig.~1 of \cite{Zhou_2023}, they compare the median value of the magnitudes from DR9-LS  and Gaia synthetic photometry. This figure shows that the zero-point calibration error traces patterns across the sky. \obiwan does not know the absolute zero-point error for the Legacy Survey, so it can not recover the trend induced by this effect. Therefore, we further use \sysnet to fix these uncorrected trends when forward modeling image systematics with \obiwan in Section~\ref{sec:obinn}.

\clearpage 
\section{Redshift-dependent systematics}
\label{sec:dndz}
Another possible source of contamination are radial systematics, caused by observing conditions that imprint a spurious variation in number density along the line of sight. Since radial systematics occur along a direction independent of angular systematics, correcting for this effect adds mitigation along a third dimension. The combination of the corrections for angular and radial systematics generates a 3D selection that accounts for non-cosmological fluctuations across the sky and along the line of sight.
In this section, we test the radial systematics that could be introduced during the target selection based on the contaminated imaging survey. 
On the other hand, independently of imaging systematics, spurious density fluctuations can also be imparted by spectroscopic systematics. These occur when incorrect or unreliable redshifts are assigned to targets and are extensively studied in \cite{KP3s3-Krolewski,KP3s4-Yu}.

\subsection{Redshift dependence simulated with Obiwan} \label{subsec:dndz_obi}
\begin{figure}
    \centering
    \includegraphics[width=\textwidth]{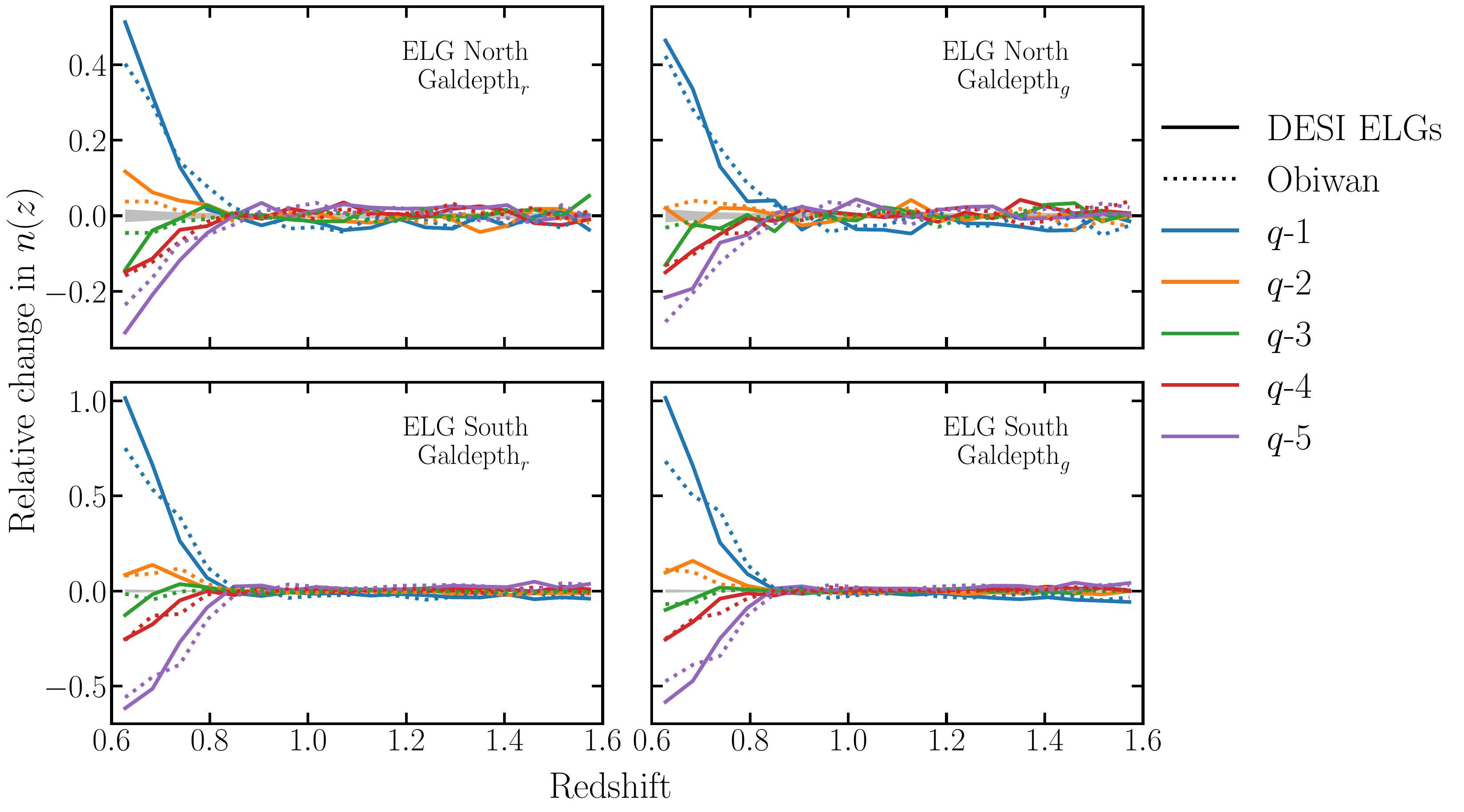}
    \caption{Variation of galaxy density as a function of redshift when splitting the sample into 5 subsamples binned by galactic depth relative to the full sample density variation. Solid lines correspond to DR1 ELG data, dotted lines show \obiwan ELG data. Top panels corresponds to North, bottom panels to South. The shadowed area represents the error for the lowest depth bin sample, i.e. the error for the blue line. \obiwan predicts imaging and redshift dependant fluctuations within statistical error over all imaging bins. In the legend, $q$-$i$ represents the $i$-th quintile.}
    \label{fig:dndz}
\end{figure}

We use simulated data from \obiwan to study how the number of galaxies within a given redshift bin, $N(z)$, is predicted to vary as a function of depth. 
Fig. \ref{fig:dndz} compares the relative change in number density of DESI ELGs to \obiwan predictions, when splitting the sample into bins of $g$ and $r$-band imaging depth.
Full details on the calculation of the plotted quantities are presented below. 
The figure shows a strong dependence of $N(z)$ on depth for $z$ near and less than 0.8, manifesting radial systematics due to the imaging attribute. The minimum redshift of $z=0.8$ adopted for DESI DR1 ELG analysis was chosen in large part due to the extreme $N(z)$ fluctuations that we find for $z<0.8$. The comparison of the prediction to the DESI ELGs (solid lines) shows a very good agreement albeit with some fluctuations. We therefore derive a model fit to Obiwan to construct a radial selection function. While the DESI DR1 analysis~\cite{DESI2024.II.KP3,DESI2024.III.KP4} does not include ELGs $z <0.8$ which exhibit strong radial dependence, in this paper, we study the effect of the radial selection function for the entire ELG redshift range ($0.6<z<1.6$) as well as the DR1 range ($0.8<z<1.6$). 

Here we describe how we measure the $N(z)$ fluctuations and estimate its error:
\begin{enumerate}
    \item Generate a set of redshift bins, in our case, we generate 10 linearly spaced redshift bins within $0.6<z<1.6$.
    
    \item Divide the whole range of the depth values, regardless of redshift, into five 20-percentile samples (i.e. quintiles).
    
    \item  Calculate the weighted $N(z)$ within each depth sample, $N(z,\mathrm{depth})$.
    
    \item Calculate the normalized relative $N(z)$ for each sample by 
    \begin{equation}
        R(z,\mathrm{depth}) = \frac{N_\mathrm{full}}{N(\mathrm{depth})} \cdot \frac{N(z,\mathrm{depth})}{N(z)},
    \end{equation}
    where $N_\mathrm{full}$ is the total number of galaxies within the $0.6<z<1.6$ redshift range, and $N(\mathrm{depth})$ is the total number of galaxies within the depth sample.
    
    \item Errors are estimated assuming Poisson statistics,
    \begin{equation}
        \sigma_{R} =  \frac{N_\mathrm{full}}{N(\mathrm{depth})} \cdot \frac{\sqrt{N(z,\mathrm{depth})}}{N(z)}.
    \end{equation}
\end{enumerate}

The match between the \obiwan predictions and the ELG data suggests that we can model the behavior found in \obiwan and apply the results to the DESI DR1 ELG data.
We find that the redshift and depth dependence can be separated into two functions that can be multiplied to obtain the total model, i.e.:
\begin{equation}
    R(z,\text{depth}) = F(z)\cdot G(\text{depth}),
\end{equation}
where $F(z)$ is a broken linear function that models the redshift dependence:
\begin{equation}
F(z) = 
\begin{cases}
    \begin{array}{lr}
        a+bz, & z < c \\
        a+bc, & z > c \\
    \end{array},
\end{cases}
\end{equation}
and $G(\text{depth})$ is an erf function that models the depth dependence:
\begin{equation}
G(\text{depth}) = d \cdot \text{erf}\left(\frac{\text{depth}}{e}+f\right) + g,
\end{equation}
where $a$, $b$, $c$, $d$, $e$, $f$ and $g$ are free parameters. 
The two models are fit simultaneously to binned \obiwan data by minimizing $\chi^2$, which gives us the optimal parameters needed for applying the model to DESI DR1 ELG data. After applying the model, we convert the result into weights that are used to mitigate depth-dependent $N(z)$ fluctuations. The weights, $w_{N(z)}$, are created by 
\begin{equation}
    w_{N(z)} = \frac{1}{1+R(z,\text{depth})},
\end{equation}
the $+1$ in the denominator is accounting for the fact that we fit to $R(z,\text{depth}) - 1$. 

The fitting method just described is useful for modeling one imaging depth band at a time. However, we can extend this procedure to fit the model to another depth bin, we do this by iteratively performing the fit. First, we fit the model in the $r$-band and obtain a set of $w_{1,N(z)}$ (subscript index is to specify which iteration the weights correspond to) for the \obiwan results and another set of weights for the data. Then we fit the model in the $g$-band; however, now the \obiwan results are weighted by the set of $w_{1,N(z)}$ weights. We use the bestfit parameters obtained from the second \obiwan fit iteration to obtain another set of $w_{2,N(z)}$ for the data, which we use to generate $w_{N(z)} = w_{1,N(z)} \times w_{2,N(z)}$ which accounts for the radial systematics due to imaging depth in the $r$ and $g$ bands.

We combine our redshift dependent selection with the angular selection function to form a three-dimensional selection function. The combination is done through their multiplication. In this work, for convenience, when studying the 3D selection function we consider the fiducial angular selection function given by the weights obtained in Section~\ref{sec:sysnet}, namely $w_\mathrm{SN}$, unless stated otherwise. The systematic weights in the $0.6<z<0.8$ redshift range are assigned based on the \elgo angular selection function. Hence, we represent the 3D imaging weights as
\begin{equation}
    w_\mathrm{SN+N(z)} = w_\mathrm{SN} \times w_\mathrm{N(z)}.
\end{equation}
We assume the addition of the radial selection function does not affect the angular distribution of targets within $0.8<z<1.6$. However, it can impact the three-dimensional clustering signal. We assess their impact on two-point clustering statistics in Section~\ref{sec:clustering_stats}.

\begin{figure}
    \centering
    \includegraphics[width=\textwidth]{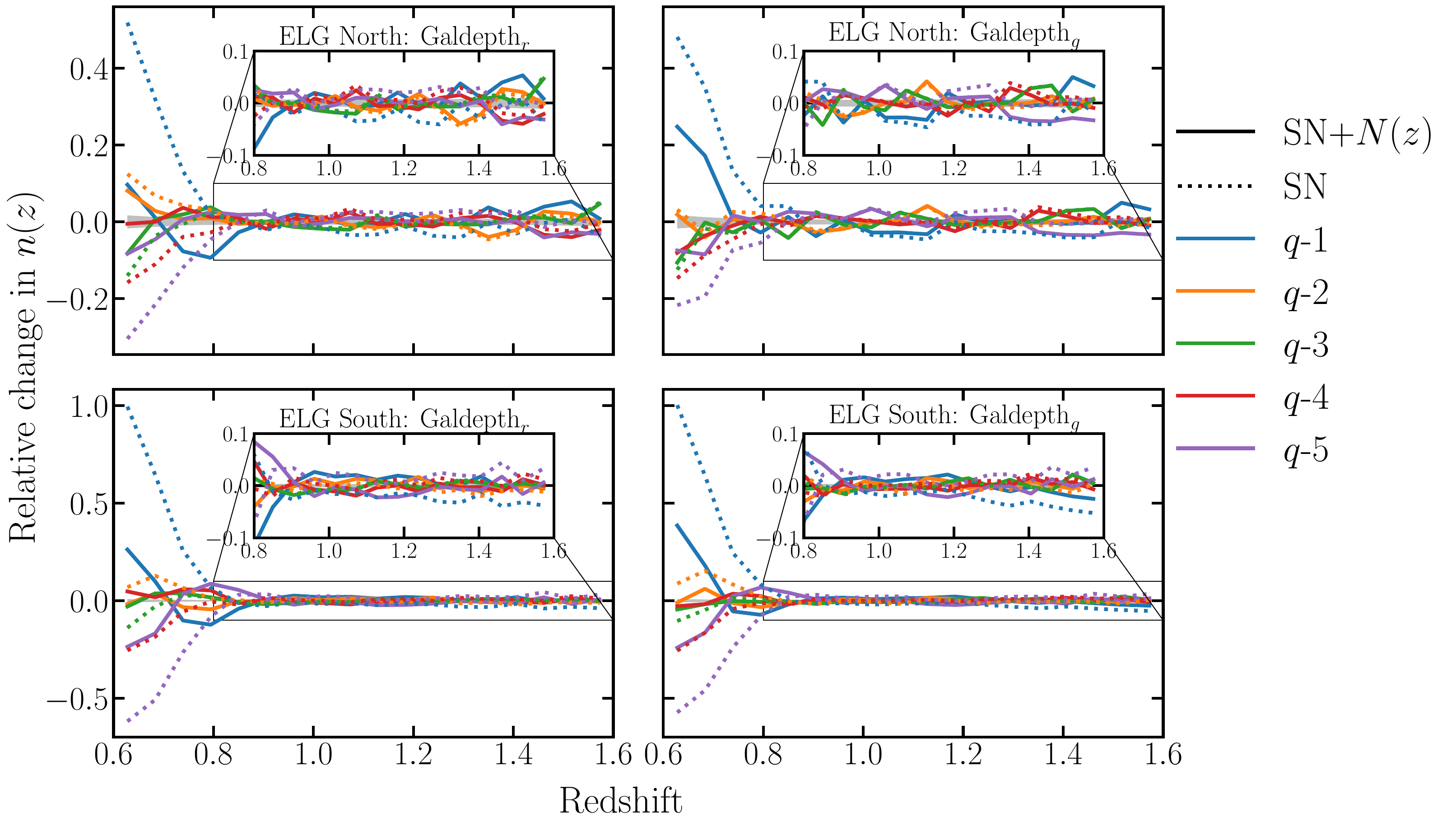}
    \caption{Relative fluctuations in galaxy density as a function of redshift when splitting the sample into 5 subsamples (into quintiles) binned by galactic depth relative to the full sample density variation. Solid lines correspond to DR1 ELG data corrected with derived $w_{N(z)}$ weights, dotted lines show unweighted DR1 ELG data (here unweighted means that it has no $w_{N(z)}$ weights applied). Top panels correspond to the North, bottom panels to the South. The shadowed area represents the error for the lowest depth bin sample, i.e. the error for the blue line. The inset zooms into the $z>0.8$ range and reduces the range of the `y' axis. The application of these radial-dependent weights reduces variations overall, especially at lower redshifts of $0.6<z<0.8$; however there are still statistically meaningful residuals. In the legend, $q$-$i$ represents the $i$-th quintile.
    }
    \label{fig:dndz-corr}
\end{figure}

In Fig \ref{fig:dndz-corr} we show the effects of using $w_{N(z)}$ weights derived from our method after iteratively fitting to galactic depth in the $r$ and $g$ bands.
Overall there is a reduction in the fluctuations. However, the low (blue lines) and high (purple lines) depth bins still are resistant to mitigation. Nonetheless, since ELGs below $z=0.8$ are not considered for DESI DR1 analysis, this level of mitigation should be enough for our assessments when looking at 2-pt statistics. It will be interesting later on to study the clustering statistics of ELGs in this low redshift bin after removing imaging and redshift-dependent variations.

\begin{figure}
    \centering
    \includegraphics[width=\textwidth]{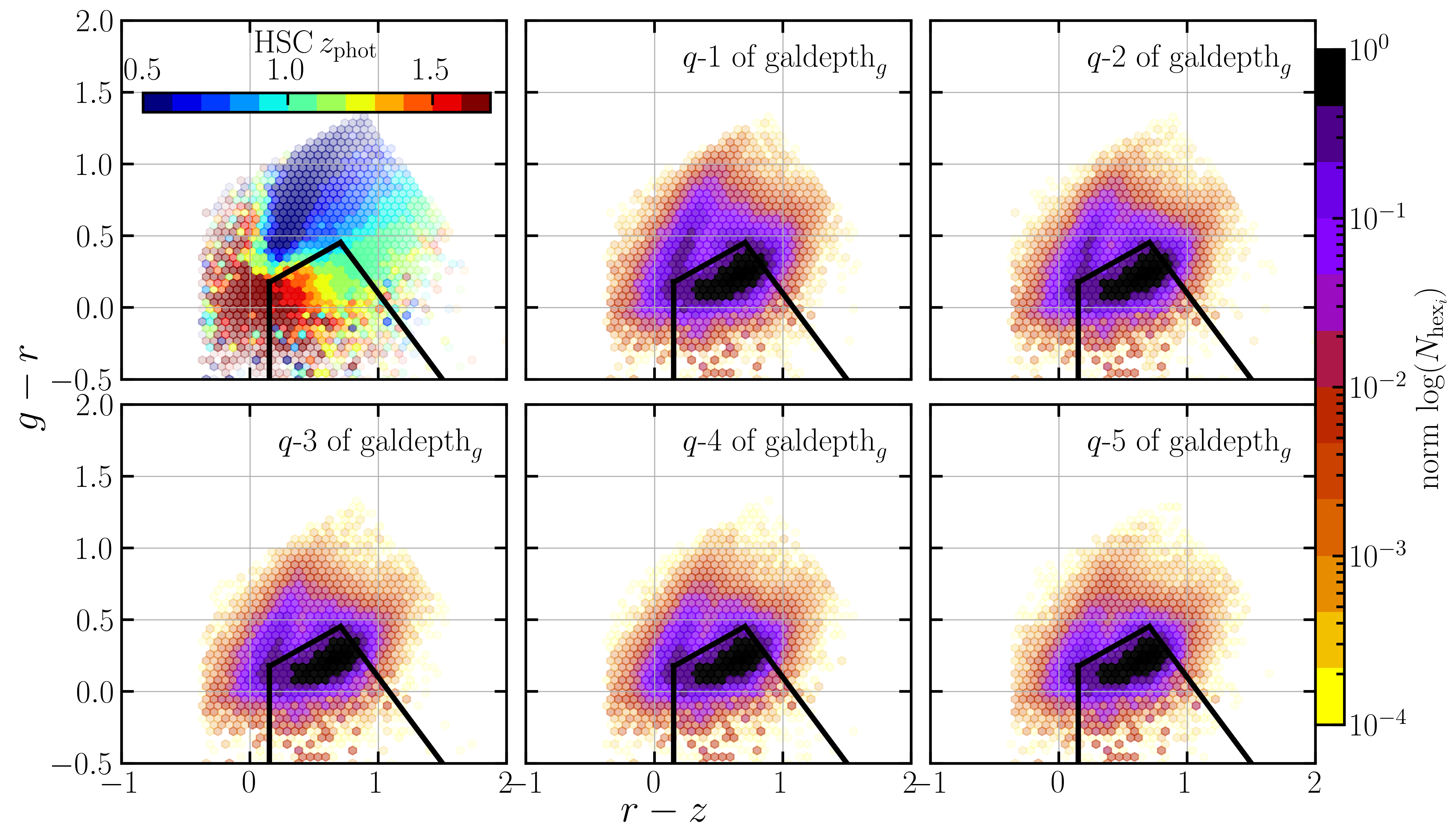}
    \caption{Color diagram for $g-r$ vs. $r-z$ made from true input photometry selected by \textsc{Obiwan}. The solid black lines show the ELG\_LOP target selection.
    The colors of the hex bins in the top-left panel show the mean HSC $z_\mathrm{phot}$, and the transparency scales with the normalized logarithm of the ELG targets selected from the true photometry.
    The colors of the bins in the rest of the panels represent the normalized logarithm of the ELG targets selected from the true photometry
    (same as the transparency scaling), with the sample in each panel cut to be within a quintile of galactic depth. 
    Notice that the axes represent the noisy observed photometry.
    In this context, the figure shows how photometric noise causes ELGs to scatter out of the selection box.
    The sample in the low galactic depth bin ($q$-1) shows higher density of targets outside the color cut selection than the high galactic depth bin ($q$-5), particularly in the region with low redshift ELGs.}
    \label{fig:color_TS}
\end{figure}

Finally, in the interest of understanding where these variations in number density originate, we use the target selection from the simulated data to select ELGs from the truth input photometry, assuming \obiwan accurately simulates the observed ELG sample.
Using a $g-r$ vs. $r-z$ diagram (see Fig.~\ref{fig:color_TS}), we illustrate the mean HSC photometric redshift ($z_\mathrm{photo}$) of the full input sample in the top-left panel. For the rest of the panels we split the sample into quintiles of galactic depth $g$ and show their density in the same color space.
From here we find that the low galactic depth bin sample has a higher density of targets in the low redshift region, which is outside the ELG color cut represented by the solid black lines. This effect is systematically increasing the number of targets with low redshift that scatter into the target selection, which is what we see in Fig~\ref{fig:dndz} where the trends in $n(z)$ for the low depth bin systematically increase as redshift decreases. 

\subsection{Testing the effect of extinction and color on redshift dependence}
\label{subsec:dndz_ebv}
\begin{figure}
    \centering  
    \includegraphics[width=\textwidth]{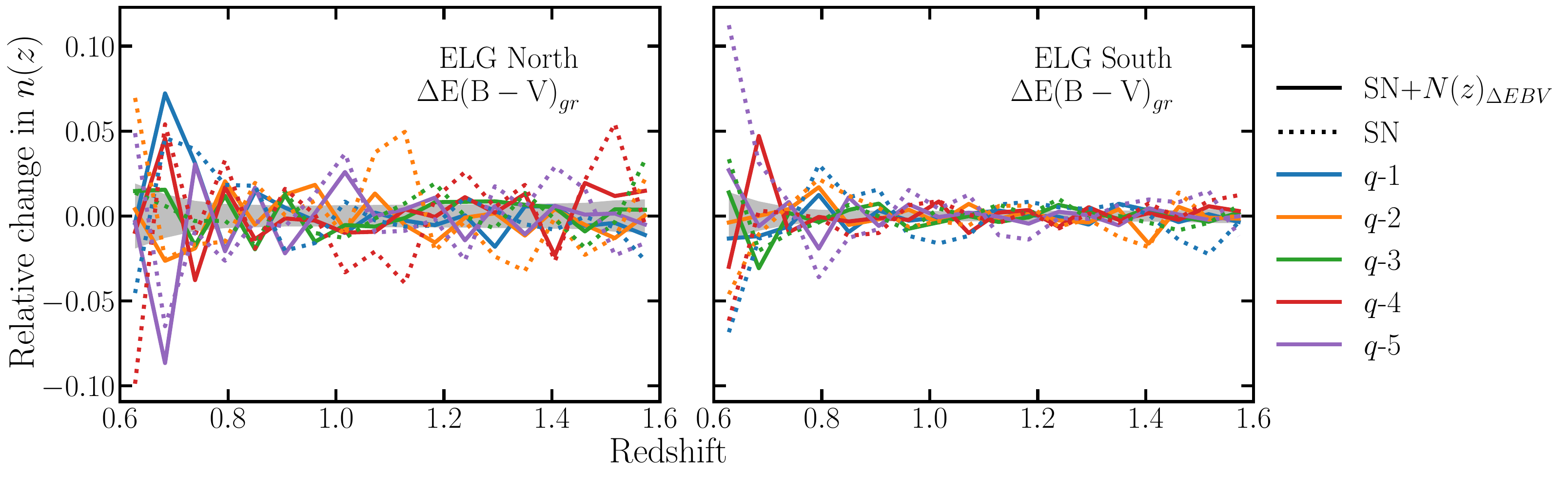}
    \caption{
    Variation of galaxy density as a function of redshift when splitting the sample into 5 subsamples (into quintiles) binned by $\Delta E(B-V)_{gr}$ relative to the full sample density variation. The left panel corresponds to the North, right panel to the South. The shadowed area represents the error for the lowest depth bin sample, i.e. the error for the blue line. The solid line represents the ELG sample with no radial systematic correction, while the dashed is with correction applied. In the legend, $q$-$i$ represents the $i$-th quintile.
    }
    \label{fig:dndz_ELGS_NS}
\end{figure}

Given that the strongest trends in the projected ELG density were found to be with the difference between galactic extinction determined using DESI stars and the SFD98 dust map (see Section \ref{sec:obiwan}), we also investigate whether there are $N(z)$ variations as a function of the \ebv differences. For this, we use the difference between the \ebv determined from $g-r$ colors and the SFD98 dust map, i.e. $\Delta E(B-V)_{gr}$. Fig.~\ref{fig:dndz_ELGS_NS} shows the number density fluctuations as a function redshift when splitting the DESI ELG sample into quintiles of $\Delta E(B-V)_{gr}$. The variations we observe at $z<0.8$ are small ($\sim 10\%$) compared to fluctuations induced by the galactic depth feature studied in the previous sub-section. 
In Section \ref{sec:clustering_stats}, we measure the impact of this trend on the 3D clustering by performing a preliminary fit on the trends and generating weights that we can apply to remove the relationship between the data and the feature. 

To determine whether these $n(z)$ variations could have a significant effect on the DR1 ELG clustering analysis, we perform aggressive modeling to remove the trends and then compare the resulting clustering measurements to the fiducial case in Section~\ref{sec:clustering_stats}. The fit is performed by first splitting the sample into three redshift bins $0.6<z<0.8$, $0.8<z<1.1$ and $1.1<z<1.6$. Then we further split each redshift bin into $20^\mathrm{th}$ percentile bins of the $\Delta E(B-V)_{gr}$ feature and fit the redshift dependence in each imaging bin using a spline interpolation. Finally, we pass the catalog redshifts through the models within each imaging bin and generate weights by taking the inverse of the model output and normalizing them to unity by their mean. 

We split the ELGs, after applying this weight, into quintiles of the $\Delta E(B-V)_{gr}$ feature and compute the number density of the new samples. As illustrated by the dotted lines in Fig~\ref{fig:dndz_ELGS_NS}, we were able to mitigate this trend to a sub-percent level for the South, while the North still shows fluctuations at $z<0.8$. Although there is likely overcorrection due to our choice of modeling, we are still interested in how much signal is removed or added to the clustering after removing this correlation. This is discussed later on in Section \ref{sec:clustering_stats}.

\section{Constructing a hybrid model of the forward modeling (\textsc{Obiwan}) and the regression method (\textsc{SYSNet})}
\label{sec:obinn}
One disadvantage of the regression-based mitigation method such as \sysnet is the potential over-correction that tends to increase when the flexibility of the mitigation method increases. Such over-corrections result in large-scale mode removal, which reduces the signal-to-noise of any clustering measurements and, if uncorrected, biases such measurements low. Estimating the effect of such overfitting typically requires applying the same mitigation method to a suite of mocks and comparing their mean clustering with and without mitigation. Such a procedure was previously applied in \cite{rezaie2023local} and for the DESI 2024 analyses as described in \cite{DESI2024.II.KP3, DESI2024.V.KP5}.
On the other hand, the forward model, such as \textsc{Obiwan}, has predictive power, free from the over-correction problem. However, its flexibility in terms of mitigation is limited to the physics that went into the model.
Ideally, we could develop a hybrid method that combines the strengths of both complementary methods. For instance, one approach could involve reducing the unnecessary flexibility in the regression-based method by substituting several input imaging attributes with predictions generated by \textsc{Obiwan}. This section represents the initial proof of concept for this hybrid approach.

In particular, \obiwans performance when predicting a number of the systematic trends, such as galactic extinction and depth, observed in the DESI DR1 ELG target density motivates us to use \obiwan results in conjunction with \sysnet for modeling imaging systematics across the sky and along the line of sight. We achieve this by the simple combination of imaging weights derived in the following subsections.

\subsection{First step: Modeling Obiwan with SYSNet}
\label{subsec:sysnet}
We first prepare a set of imaging weights generated by modeling the ELG density simulated by \obiwan with \textsc{SYSNet}. The simulated data is prepared similarly as the real ELG data; the key differences are that we do not need to weight the simulated data, and the pixel completeness is determined by the ratio of the number of \obiwan inputs per pixel over the mean number of \obiwan inputs per pixel.
Then \sysnet is run on the simulated data, and imaging weights are produced in the same manner described for Eq. \ref{eq:wsys}. The imaging weights are normalized by their mean value and then projected to a \textsc{HEALPix} map. The \obiwan based imaging weights derived from \textsc{SYSNet}, $w_{\mathrm{obi}}$, are assigned to objects by converting catalog RA and Dec to \textsc{HEALPix} and using these pixel values for $w_{\mathrm{obi}}$ assignment. As we do not need to worry about overcorrection when fitting the simulated \obiwan data, we utilize the same imaging attributes used in Section~\ref{subsec:sysnet_elgs}.

Again, these \obiwan based photometric weights do not mitigate all spurious correlations between data and systematics to the level needed for DESI DR1 ELGs; this is illustrated in Fig. \ref{fig:2nd_step}. The figure shows the relative \elgt target density as a function of the stellar density for different mitigation weights. 
The black markers represent the ELG density weighted using DESI fiducial imaging weights which are \sysnet based, $w_\mathrm{SN}$ (denoted as `SN'); 
the dot-dashed represents the case where we mitigate only the systematics that are modeled by \obiwan, with $w_\mathrm{obi}$ (denoted as `obi');
dotted blue curves have no mitigation applied. Qualitatively, we can see that the \obiwan based weights alone do not mitigate the trends to the same level as the fiducial weights (weights derived from \sysnet only). Specifically, in the North, we observe the significant residual trend for the regions with stellar density less than $11,000 \, \mathrm{deg}^{-2}$, while for the South, we observe a substantial residual over the entire range of the stellar density. This observation is further corroborated by the $\chi^2$ values, which let us quantitatively assess how much of the systematic trend was removed through the application of the imaging weights when compared to the null trend. For the fiducial weights, we obtain $\chi^2$ values of 5.4 and 4.4, for the North and South regions respectively, which are close to the degrees of freedom (10 $dof$) and significantly lower than the 73.7 and 107.7 obtained from the \obiwan based weights.

\subsection{Second step: Model the residual with SYSNet}
\label{subsec:second_step}
Given that there are still statistically significant imaging trends in the ELG sample after up-weighting the target density field with imaging weights derived from \textsc{Obiwan}, we use \sysnet again to mitigate the residual systematics. The \sysnet weight derived for this \obiwan residual we denote as  $w_\mathrm{SN,res}$. 
For this iteration of the weights, the data used for training has $w_\mathrm{obi}$ applied and then we train the data in the same manner as the DESI fiducial \sysnet based weights for ELGs, also using the same set of imaging maps (see Section \ref{sec:sysnet}).
Then we combine the weights by simply multiplying them by the \obiwan based weights, which generates the combined \obisys weights which we denote as $w_\mathrm{obiNN}$, explicitly
\begin{equation}
    w_\mathrm{obiNN} = w_\mathrm{SN,res} \times w_\mathrm{obi}.
\end{equation}

\begin{figure}
    \centering
    \includegraphics[width=0.9\textwidth]{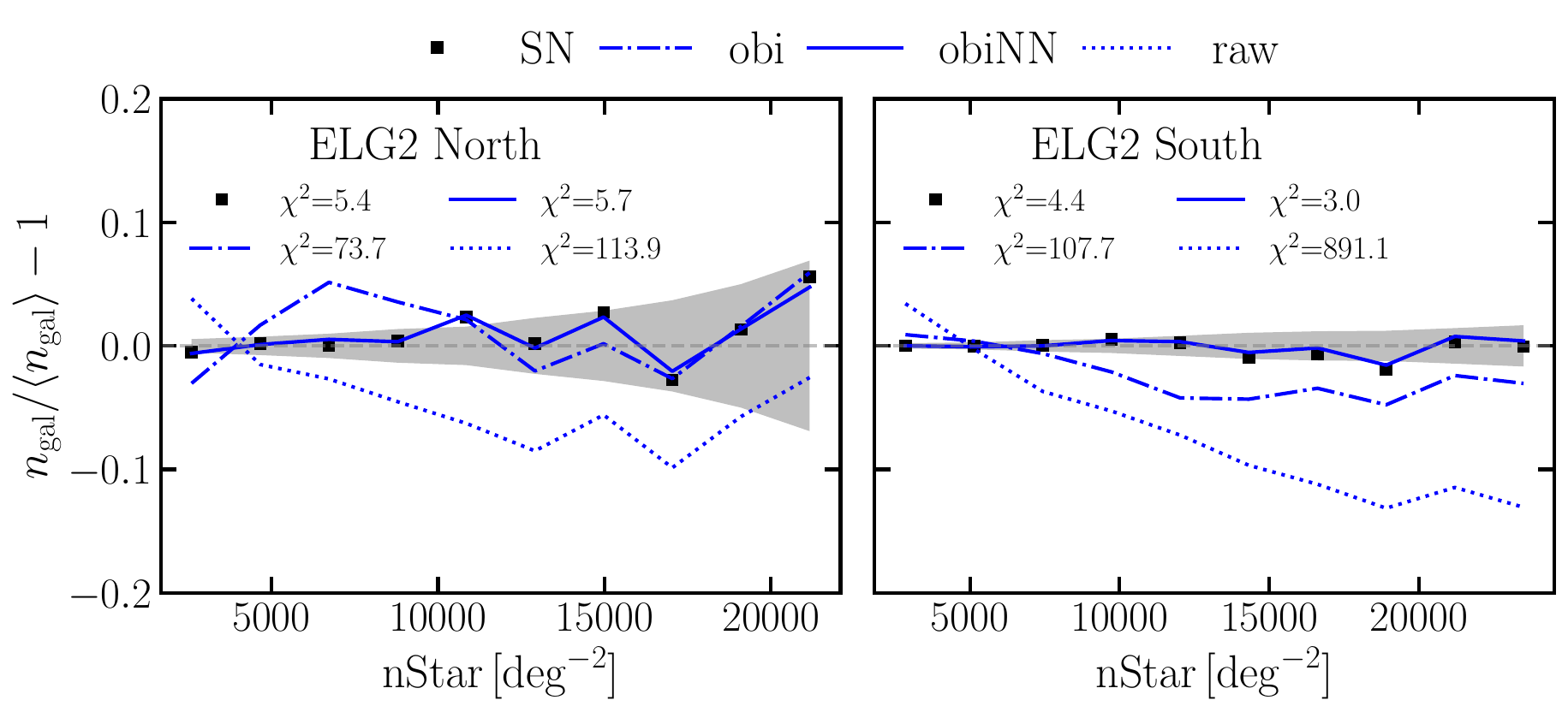}
    \caption{Galaxy mean density of the \elgt sample as a function of stellar density. The black markers correspond to the ELG density weighted using DESI fiducial imaging weights ($w_\mathrm{SN}$), the dashed-dotted blue line is weighted using $w_\mathrm{obi}$, while the solid blue curve is weighted using the combined $w_\mathrm{obiNN}$, and dotted blue has no imaging weights applied. $\chi^2$ values are computed for each weight choice. The left panel shows results for the North \elgt and the right panel for the South \elgt. \obiwan based imaging weights alone are not enough to mitigate to the same degree as the fiducial weights. We try to remove residual contamination with $w_\mathrm{obiNN}$ weights which show a degree of mitigation comparable to that of the fiducial weights.}
    \label{fig:2nd_step}
\end{figure} 

The effects of these new imaging weights based on the combination of forward modeling and regression techniques are shown in Fig. \ref{fig:2nd_step}. Directing our attention to the solid blue line, which represents the ELG mean galaxy density weighted by $w_\mathrm{obiNN}$, 
we observe a major reduction in the the spurious correlation between stellar density and ELG mean density, for the North and the South \elgt samples, compared to the case when just $w_{\rm obi}$ is applied.
We find that, as expected, $w_\mathrm{obiNN}$ can remove \obiwan residuals and the performance of $w_\mathrm{obiNN}$ and the default, $w_\mathrm{SN}$, agree. In detail, we find $\chi^2$ of $w_\mathrm{obiNN}$ are slightly lower than that of $w_\mathrm{SN}$.
For the case of stellar density shown in \cref{fig:2nd_step}, for $w_\mathrm{obiNN}$ we find $\chi^2$ values of 5.7 and 3.0, for the North and the South \elgt respectively, while for $w_\mathrm{SN}$ we find 5.4 and 4.4.

\begin{figure}
    \centering
    \includegraphics[width=\textwidth]{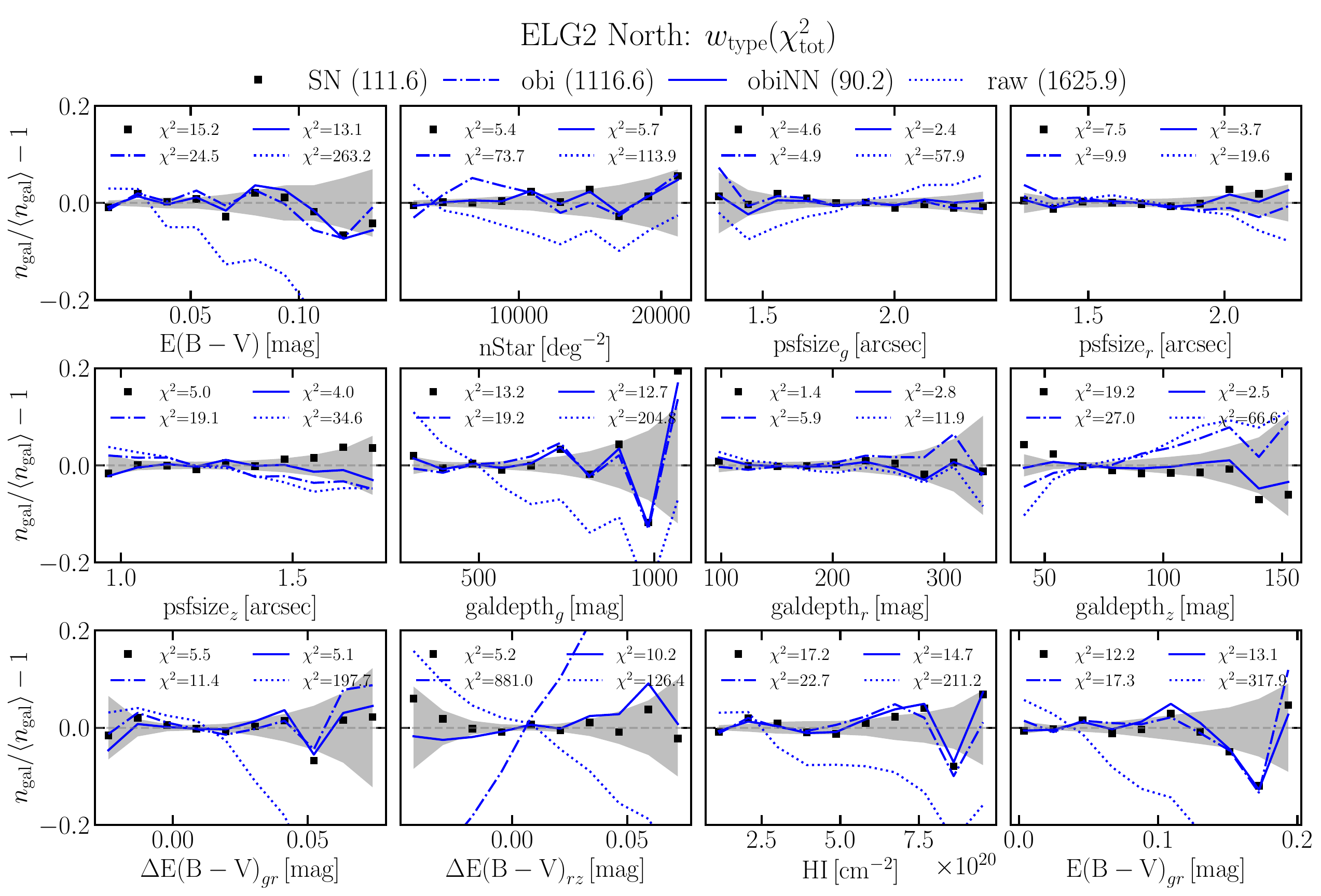}
    \caption{Galaxy mean density of DESI DR1 ELGs as a function of several imaging systematics. The black markers correspond to the ELG density weighted using DESI fiducial imaging weights ($w_\mathrm{SN}$), dashed-dotted blue line is weighted using $w_\mathrm{obi}$, solid blue line is weighted using $w_\mathrm{obiNN}$,
    dotted blue has no imaging weights applied. $\chi^2$ values are computed for each weight choice per imaging feature, and the total $\chi^2_\mathrm{tot}$ is shown in the title. Each panel has $dof=10$, for a total of $dof=120$.}
    \label{fig:elg-dens_N}
\end{figure}

\begin{figure}
    \centering
    \includegraphics[width=\textwidth]{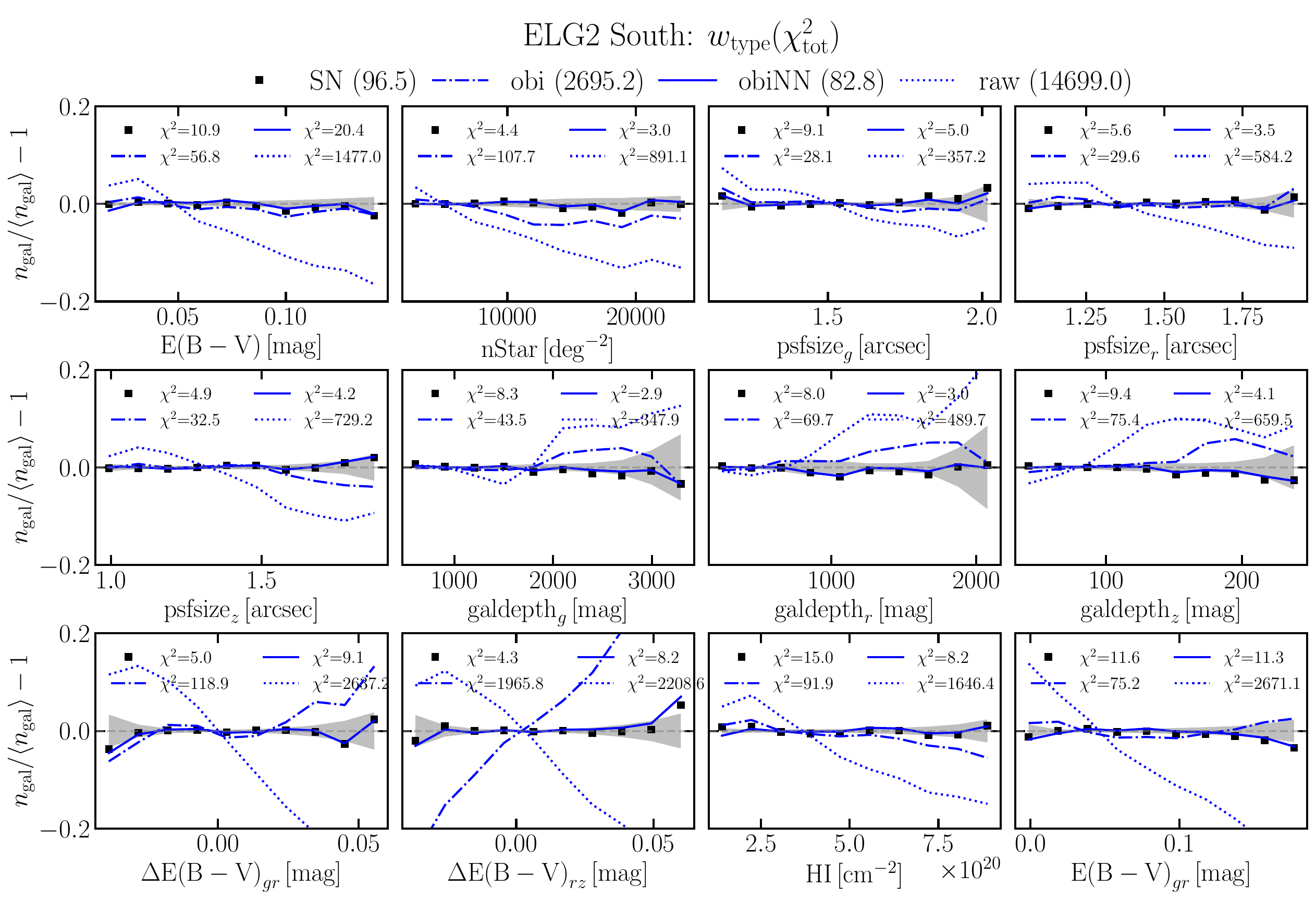}
    \caption{Same as figure \ref{fig:elg-dens_N}, but for the \elgt South sample.}
    \label{fig:elg-dens_S}
\end{figure}

In Figs. \ref{fig:elg-dens_N} and \ref{fig:elg-dens_S}, we show the mean galaxy density of \elgt as a function of an extended set of imaging attributes. 
Under the titles of the figures, we show the total $\chi^2$ ($dof=120$) obtained when summing over all the $\chi^2$ for each imaging attribute. 
For the North ELGs, obiNN weights have a smaller $\chi^2=90.2$ compared to the fiducial case of $\chi^2=111.6$. 
For the South sample, the obiNN results yield $\chi^2=82.8$ compared to the fiducial case of $\chi^2=96.5$. 
The reduction in total $\chi^2$ values suggests that the obiNN weights are removing a small amount of variance aligned with our potential systematic templates from the ELG density field. 
The \elgo sample exhibits similar behavior and rather than include plots for it, we summarize the $\chi^2$ results as: $\chi^2=92.6$ (obiNN) and $\chi^2=127.0$ (fiducial) for the \elgo North, $\chi^2=82.6$ (obiNN) and $\chi^2=81.7$ (fiducial) for the \elgo South.
In Section \ref{sec:clustering_stats}, we will compare our hybrid approach to the fiducial scheme in terms of the resulting clustering signal.

\clearpage \section{Impact on clustering statistics and BAO measurements}\label{sec:clustering_stats}
\subsection{Impact of imaging systematics on clustering statistics}
\label{subsec:elg_clustering}
Here we discuss the impact of different mitigation schemes on clustering statistics, the two-point correlation function and the power spectrum. To facilitate the comparison, we take the fiducial scheme as the baseline and show the fractional errors of the differences between the mitigation scheme and the fiducial weighting scheme as
\begin{equation}\label{eq:fracsigma}
    \frac{\nu^\prime_\ell - \nu_\ell}{\sigma_{\nu_\ell}},
\end{equation}
where $\nu_\ell$ is the clustering statistics multipole after up-weighting galaxies with the fiducial scheme, while $\nu^\prime_\ell$ is when weighting with a particular weighting scheme, and $\sigma_{\nu_\ell}$ corresponds to the error associated with the fiducial scheme.
Here $\sigma_{\nu_\ell}$ is taken from the analytic covariance matrices discussed in \cite{DESI2024.III.KP4}.

\begin{figure}
    \centering  
    \includegraphics[width=\textwidth]{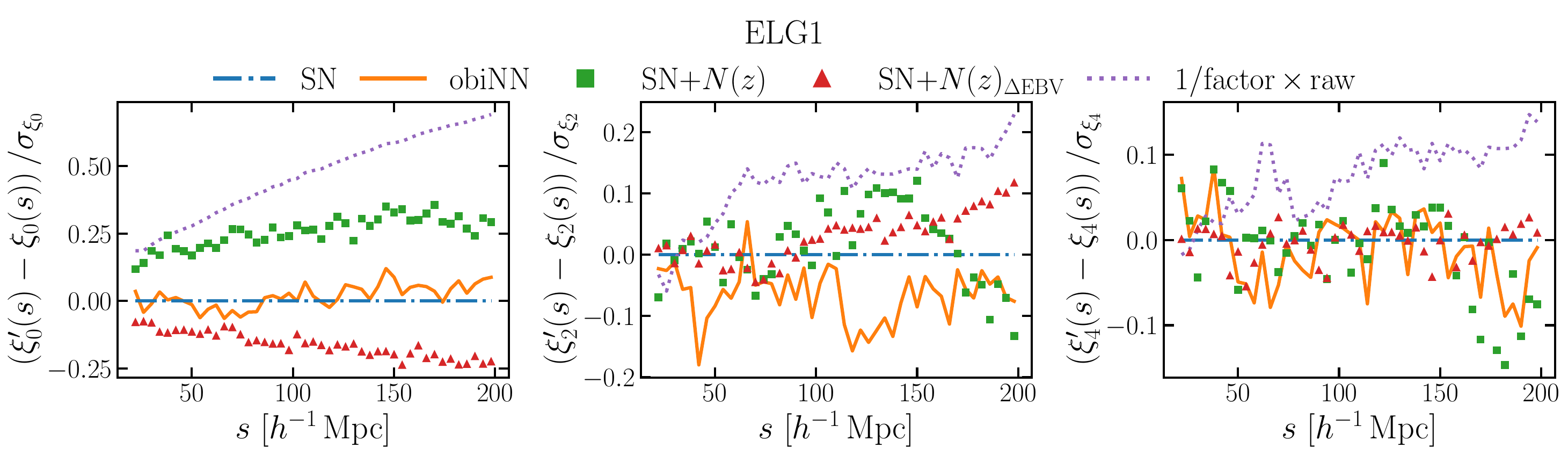}
    \includegraphics[width=\textwidth]{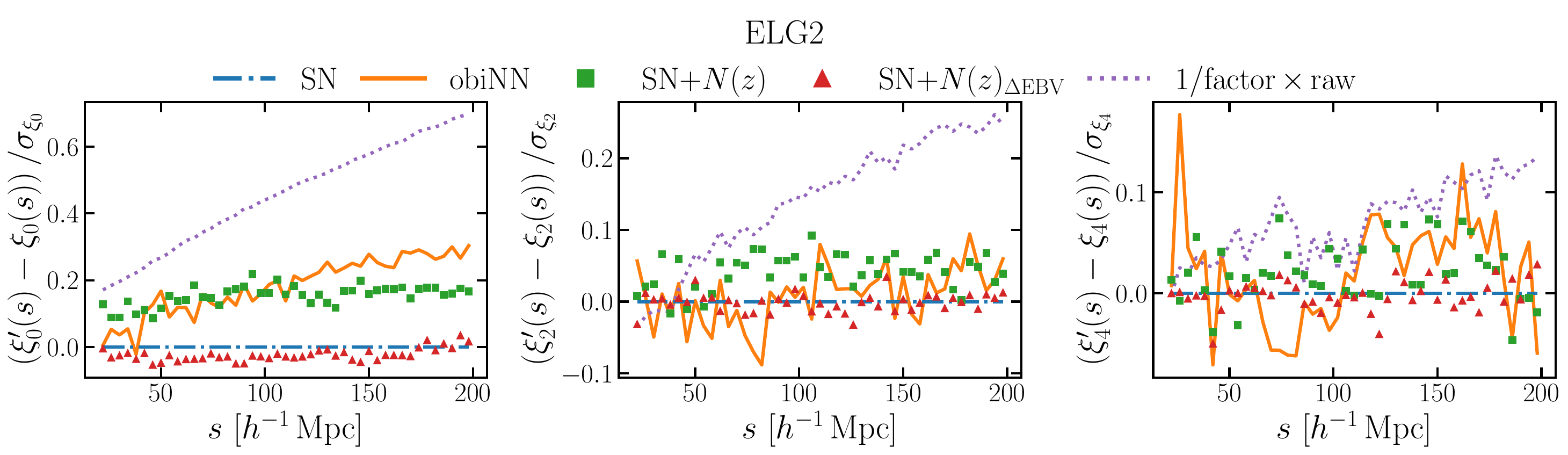}
    \caption{Fractional errors of the differences between the correlation function multipoles for the weighted $\xi^\prime_\ell(s)$ and fiducial $\xi_\ell(s)$ for the combined ELG sample (NGC+SGC). From left to right, the figure shows the monopole, quadrupole, and hexadecapole. The top panels correspond to \elgo, while the bottom panels are for \elgt. 
    The dotted dashed blue line corresponds to DESI fiducial weights,
    solid orange lines for our combined \obiwan and \sysnet weights,
    the green squares show data weighted with fiducial weights and $N(z)$ weights,
    the red triangles show data weighted with fiducial weights and $N(z)_\mathrm{\Delta EBV}$ weights,
    dotted purple lines show when no imaging weights are applied. 
    To facilitate comparison, the monopole of the `raw' signal is divided by a factor of 50, and the rest of the multipoles are divided by a factor of 10. 
    Notice that no BAO-like feature is observable in these plots.}
    \label{fig:xi_sys_ELGS_NS}
\end{figure}

\begin{figure}
    \centering  
    \includegraphics[width=\textwidth]{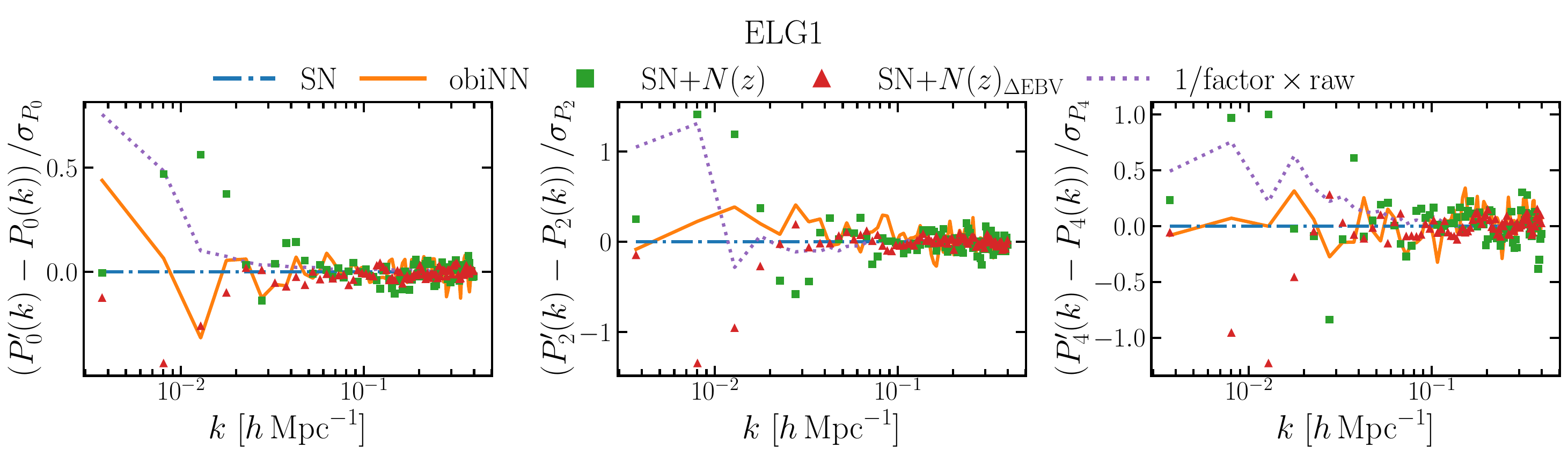}
    \includegraphics[width=\textwidth]{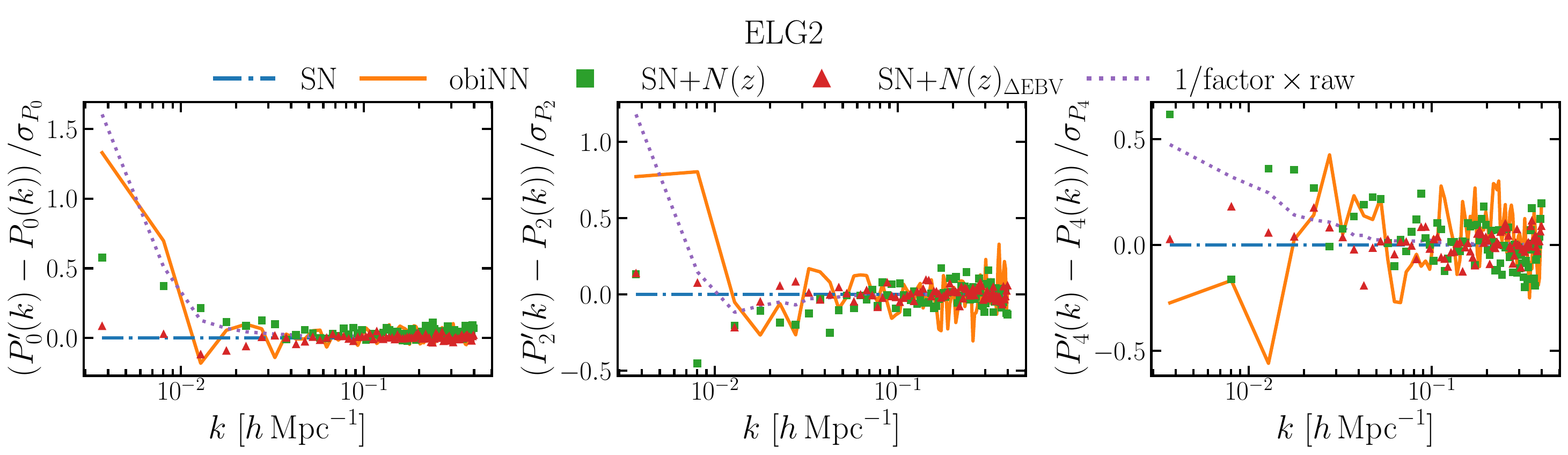}
    \caption{Fractional errors of the differences between the power spectrum multipoles for the weighted $P^\prime_\ell(k)$ and fiducial $P_\ell(k)$ for the combined ELG sample (NGC+SGC). From left to right, the figure shows the monopole, quadrupole, and hexadecapole. The top panels correspond to \elgo, while the bottom panels are for \elgt. 
    The dotted dashed blue line corresponds to DESI fiducial weights,
    solid orange lines for our combined \obiwan and \sysnet weights, 
    the green squares show data weighted with fiducial weights and $N(z)$ weights,
    the red triangles show data weighted with fiducial weights and $N(z)_\mathrm{\Delta EBV}$ weights,
    dotted purple lines show when no imaging weights are applied. To facilitate comparison of the \elgo sample, the monopole of the `raw' signal is divided by a factor of 100, and the rest of the multipoles are divided by a factor of 35. For the \elgt sample, all `raw' signal multipoles are divided by a factor of 100.
    Similar to the 2-pt correlation function results, we observe no BAO-like feature in these power spectra.
    }
    \label{fig:pk_sys_ELGS_NS}
\end{figure}

In Fig. \ref{fig:xi_sys_ELGS_NS} we show the fractional errors, as in Eq. \ref{eq:fracsigma}, from comparing the two-point correlation function multipoles for a particular weighting scheme, $\xiellprime$, to the multipoles from the fiducial scheme $\xiell$. Fig~\ref{fig:pk_sys_ELGS_NS} is the Fourier space version of Fig~\ref{fig:xi_sys_ELGS_NS}. 
The 2-pt correlation function is measured in $4\,\mpch$ bins from $s=20$ to $s=200\,\hmpc$ (45 $s$-bins), while the power spectrum in  $5\,\hmpc$ bins from $k=0.005$ to $k=0.4\,\hmpc$ (80 $k$-bins). The $\chi^2$ statistic we compute is over these bins.
Each different line-style and color combination denotes a choice of weights when computing the correlation function, and to facilitate the comparison between weighting schemes the `raw' (without imaging weights) clustering signal is divided by some constant factor specified in the corresponding figure caption.

Using the fiducial weighting for DR1 ELGs as a baseline, we determine the $\chi^2$ values by comparing the multipole amplitudes between mitigation schemes and the baseline. In Table~\ref{tab:chi2} we present the $\chi^2$ values obtained for the \elgo and \elgt samples (NGC+SGC). Furthermore, the columns containing the $\chi^2$ values for the multipoles also show the cumulative $\chi^2$ up to the corresponding multipole. 
The cumulative $\chi^2$ values we present include the cross-covariance between the multipoles and these are particularly useful for quantifying the total variance imparted by the weighting scheme and provide a total quantification of the fractional errors presented in Figs.~\ref{fig:xi_sys_ELGS_NS} and \ref{fig:pk_sys_ELGS_NS}. Further, taking their square root value, namely $\sqrt{\chi^2}$, provides the upper limit on how much a single derived parameter could change in terms of N$\sigma$ (i.e., it is the significance in the case where the parameter causes the model to change in exactly the same way as the change in weighting shifts the data vector).
Below we discuss our findings for each weighting scheme, except for the blue dotted-dashed lines as they represent the fiducial case, which is the baseline and should be null.

\begin{table}
    \centering
\caption{$\chi^2$ values obtained from comparing the correlation function and power spectrum multipoles computed using DESI fiducial weights compared to the different weight choices. The upper half of the table corresponds to the \elgo sample and shows $\chi^2$ values for the combination of both regions NGC+SGC, while the lower half is for the \elgt sample. Inside the parenthesis, we show the cumulative $\chi^2$ up to the given multipole.}
\label{tab:chi2}
    \begin{tabular}{lllll}
    \hline
    \textbf{Redshift}  & \textbf{Weights}    & $\ell=0$   & $\ell=2$   & $\ell=4$   \\
    \hline
    \hline
 $\mathbf{0.8<z<1.1}$ & \textbf{$\xi_\ell(s)$ GCcomb}                &            &                   &                  \\
                      & obiNN                          & 0.082      & 0.139 (0.240)     & 0.086 (0.345)    \\
                      & SN+$N(z)$                      & 0.383      & 0.126 (0.514)     & 0.119 (0.623)    \\
                      & SN+$N(z)_{\Delta\mathrm{EBV}}$ & 0.161      & 0.036 (0.206)     & 0.024 (0.231)    \\
                      & raw                            & 2969.660   & 13.649 (2986.846) & 7.609 (3011.485) \\
 $\mathbf{0.8<z<1.1}$ & \textbf{$P_\ell(k)$ GCcomb}                &            &                      &                     \\
                      & obiNN                          & 0.567      & 0.287 (1.168)        & 0.204 (1.506)       \\
                      & SN+$N(z)$                      & 0.632      & 0.856 (1.352)        & 0.595 (2.044)       \\
                      & SN+$N(z)_{\Delta\mathrm{EBV}}$ & 0.263      & 0.574 (0.922)        & 0.338 (1.254)       \\
                      & raw                            & 5930.967   & 1512.559 (8082.391)  & 205.631 (10124.069) \\
 \hline
 $\mathbf{1.1<z<1.6}$ & \textbf{$\xi_\ell(s)$ GCcomb}                &            &                   &                  \\
                      & obiNN                          & 0.294      & 0.077 (0.378)     & 0.108 (0.490)    \\
                      & SN+$N(z)$                      & 0.165      & 0.045 (0.235)     & 0.049 (0.293)    \\
                      & SN+$N(z)_{\Delta\mathrm{EBV}}$ & 0.016      & 0.011 (0.029)     & 0.012 (0.041)    \\
                      & raw                            & 3460.312   & 20.142 (3539.575) & 8.266 (3566.301) \\
$\mathbf{1.1<z<1.6}$ & \textbf{$P_\ell(k)$ GCcomb}                &            &                      &                     \\
                      & obiNN                          & 2.140      & 0.607 (2.609)        & 0.222 (2.947)       \\
                      & SN+$N(z)$                      & 0.525      & 0.196 (0.748)        & 0.229 (1.034)       \\
                      & SN+$N(z)_{\Delta\mathrm{EBV}}$ & 0.053      & 0.046 (0.092)        & 0.037 (0.137)       \\
                      & raw                            & 25982.909  & 8145.849 (26967.691) & 448.933 (27364.509) \\
\hline
    \hline
    \end{tabular}
\end{table}

First, let us consider the case of applying the combined \obisys weights (obiNN; solid orange). Starting from Fig.~\ref{fig:xi_sys_ELGS_NS}, we notice that the \elgo monopole starts to scatter around the baseline with a slight positive offset that increases up to $\sim0.1\sigma$ at large scale. The \elgo quadrupole also shows scatter but around $\sim-0.8\sigma$, while the hexadecapole scatter the baseline over all scales. 
On the other hand, we find that the \elgt monopole increases more rapidly from $\sim0$ to $0.3\sigma$.
Even with these discrepancies, the $\chi^2$ that we obtain with this weighting scheme is small, corresponding to $\sqrt{\chi^2}=\sqrt{0.345}=0.59\sigma$ and $\sqrt{\chi^2}=\sqrt{0.490}=0.7\sigma$ for \elgo and \elgt respectively, when compared to the baseline. We expect this since obiNN weights are obtained by training on the same set of imaging maps used for obtaining the fiducial weights. In Fourier space, as shown in Fig.~\ref{fig:pk_sys_ELGS_NS}, at the larger scales we find the largest deviations from the baseline, across both ELG samples. However, these fluctuations are greater for the \elgt sample, as we already saw in the configuration space, with $\chi^2$ values corresponding to $\sqrt{\chi^2}=\sqrt{2.947}=1.72\sigma$ compared to the $\sqrt{\chi^2}=\sqrt{1.506}=1.23\sigma$ for the \elgo sample.

In the case of combining the fiducial \sysnet weights with depth-dependent $N(z)$ weights (SN+$N(z)$; green squares), in Fig.~\ref{fig:xi_sys_ELGS_NS} we observe a positive offset that slightly increases with separation in the monopole for both samples, the offset for the \elgo monopole is between $\sim0.1$ and $0.3\sigma$, while $\sim0.1$ and $0.2\sigma$ for \elgt.
For the higher-order multipoles, \elgo shows scatter around the baseline, however, the \elgt quadrupole shows a positive offset around $\sim0.05\sigma$. From the cumulative $\chi^2$ values shown in Table~\ref{tab:chi2}, we find that $\sqrt{\chi^2}=\sqrt{0.623}=0.79\sigma$ and $\sqrt{\chi^2}=\sqrt{0.293}=0.54\sigma$ for the \elgo and \elgt samples respectively, which suggest that the total variance from adding the $\mathrm{SN}+N(z)$ weights is less than the uncertainty attributed to the baseline clustering measurement.

In Fig.~\ref{fig:pk_sys_ELGS_NS}, we see that the effect of the $\mathrm{SN}+N(z)$ weights on the power spectrum is most significant at large scales, i.e. low $k$, across both samples and their multipoles. As supported by the cumulative $\chi^2$ values which yield $\sqrt{\chi^2}=\sqrt{2.044}=1.42\sigma$ and $\sqrt{\chi^2}=\sqrt{1.034}=1.02\sigma$ for the \elgo and \elgt samples respectively, the impact of $\mathrm{SN}+N(z)$ weights is greater in Fourier space. Also, for the \elgo sample, the fluctuations around the baseline at large $k$, for the higher order multipoles in particular, are larger than those observed for the obiNN case.

Next we study the effect of the combined fiducial \sysnet weights with $\Delta \mathrm{EBV}$-dependent $N(z)$ weights ($\mathrm{SN}+N(z)_\mathrm{\Delta EBV}$; red triangles) from section \ref{subsec:dndz_ebv}. In Fig~\ref{fig:xi_sys_ELGS_NS}, where we present the 2-pt clustering in configuration space, in the \elgo sample we see that the monopole signal is lower than the baseline clustering amplitude and decreases (down to $-0.25\sigma$) as the separation distance increases. This reduction in the monopole amplitude is expected given how aggressive the treatment was (i.e., we expect it to remove some true clustering modes). For the \elgt sample, the monopole amplitude also is lower than the baseline, however, the offset is less ($\sim0.1\sigma$), and around $\sim175\,\mpch$ the amplitude fluctuates around the baseline. For the quadrupole of the \elgo sample, the amplitude fluctuates around the baseline up to $\sim100\,\mpch$ where it starts to increase up to $0.1\sigma$. For the rest of the multipoles of the \elgo and the \elgt samples, their amplitudes fluctuate around the baseline. From the cumulative $\chi^2$ values shown in Table~\ref{tab:chi2}, we find $\sqrt{\chi^2}=\sqrt{0.231}=0.48\sigma$ and $\sqrt{\chi^2}=\sqrt{0.041}=0.2\sigma$ for the \elgo and \elgt samples respectively, which show less of an impact on the clustering signal than the $\mathrm{SN}+N(z)$ weights.

Now let us see the effect of applying $\mathrm{SN}+N(z)_\mathrm{\Delta EBV}$ weights on the clustering amplitude in Fourier space, as shown in Fig.~\ref{fig:pk_sys_ELGS_NS}. For the \elgo sample, the clustering amplitude appears consistent across all multipoles until it reaches low $k$ values, where the sample is more prone to imaging systematics. At these scales, the fractional errors reach a maximum of around $\sim-0.1\sigma$, as seen for the correlation function, this seems to suggest mode removal from over-fitting in the \elgo sample. On the other hand, the multipole amplitudes for the \elgt sample fluctuate around the baseline clustering amplitude with the fluctuation being marginally larger for low values of $k$. From the cumulative $\chi^2$ values, we find $\sqrt{\chi^2}=\sqrt{1.254}=1.12\sigma$ and $\sqrt{\chi^2}=\sqrt{0.137}=0.37\sigma$ for the \elgo and \elgt samples respectively. We notice that the \elgo has a higher total variance, but it is still consistent with the baseline.

In summary, we find that changing the weighting is more significant in Fourier space than in configuration space, but when considering alternative weightings from the default case, the greatest changes are order 1. These changes can be compared to the unweighted vs. fiducial weighted case, which is order 10$^{4}$ in terms of $\chi^2$. To account for imaging systematics in \cite{DESI2024.V.KP5}, a free amplitude applied to a smooth model fit to the difference between the weighted and unweighted clustering data is applied as an extra model term. Such treatment should generously allow for the changes we observe between different weighting choices. The simplest case of the NN weights was thus chosen for the fiducial DESI DR1 ELG catalogs \cite{DESI2024.II.KP3}. However, we will find that the application of a radial selection function is necessary for ELGs between $0.6<z<0.8$, as presented in \cref{subsec:elg0_clustering}.

Finally, we discuss the unmitigated clustering signal (raw; dotted purple), which as expected shows a strong level of excess clustering.
In Fig.~\ref{fig:xi_sys_ELGS_NS}, we observe across both samples and all multipoles a significant positive offset, which increases with scale. This is expected since contamination due to imaging systematics is stronger at larger scales. From Table~\ref{tab:chi2}, we obtain $\sqrt{\chi^2}=\sqrt{3,011}=54.87\sigma$ and $\sqrt{\chi^2}=\sqrt{3,566}=59.72\sigma$ for the \elgo and \elgt samples respectively, which quantify how strong these systematics are and the importance of mitigating them. For the power spectrum (see Fig.~\ref{fig:pk_sys_ELGS_NS}), as expected, we also find strong clustering at large scales in the form of large excess clustering amplitude at low values of $k$ (as $k\rightarrow0$). The magnitude of the excess clustering observed in the power spectra is even greater than that observed in the 2-pt correlation function, as shown by the $\chi^2$ values corresponding to $\sqrt{\chi^2}=\sqrt{10,124}=100.62\sigma$ (\texttt{ELG1}) and $\sqrt{\chi^2}=\sqrt{27,364}=165.4\sigma$ (\texttt{ELG2}).

\subsection{Impact of radial systematics on \texttt{ELG0} clustering}
\label{subsec:elg0_clustering}
In this subsection, we discuss our findings on the inclusion of SN+$N(z)$ weights in the \elgz sample. As discussed in \ref{subsec:dndz_obi}, DR1 ELGs within the $0.6<z<0.8$ redshift range showed strong $N(z)$ depth-dependent variations which we corrected for by modeling these trends with \obiwan synthetic galaxies. 

Before looking at the clustering statistics of the sample, we explain some changes in nomenclature that only apply to this subsection. 
Since the DR1 ELG catalogs do not contain imaging weights for the $0.6<z<0.8$ redshift range, we generated them by using the same data preparation and \sysnet settings as the fiducial weights used for $0.8<z<1.6$. We denote these as $\mathrm{SN}_\mathrm{fid}$. 
However, we found that if we want to use $N(z)$ weights in this redshift range, we can no longer assume they are independent of angular systematics and simply multiply the angular imaging weights with the radial imaging weights as done before.
By doing so, the multiplication of both selection functions increased the clustering amplitude considerably. This happens because the radial imaging systematics for this redshift range are correlated with the angular imaging systematic. 
To handle this issue, we trained the \sysnet NN on data up-weighted by $N(z)$ weights and proceeded to train as normal. In this subsection, we denote the weights obtained through this method as $\mathrm{SN}+N(z)$.
Finally, the clustering signal with no weights applied remains as `raw'.

\begin{figure}
    \centering  
    \includegraphics[width=\textwidth]{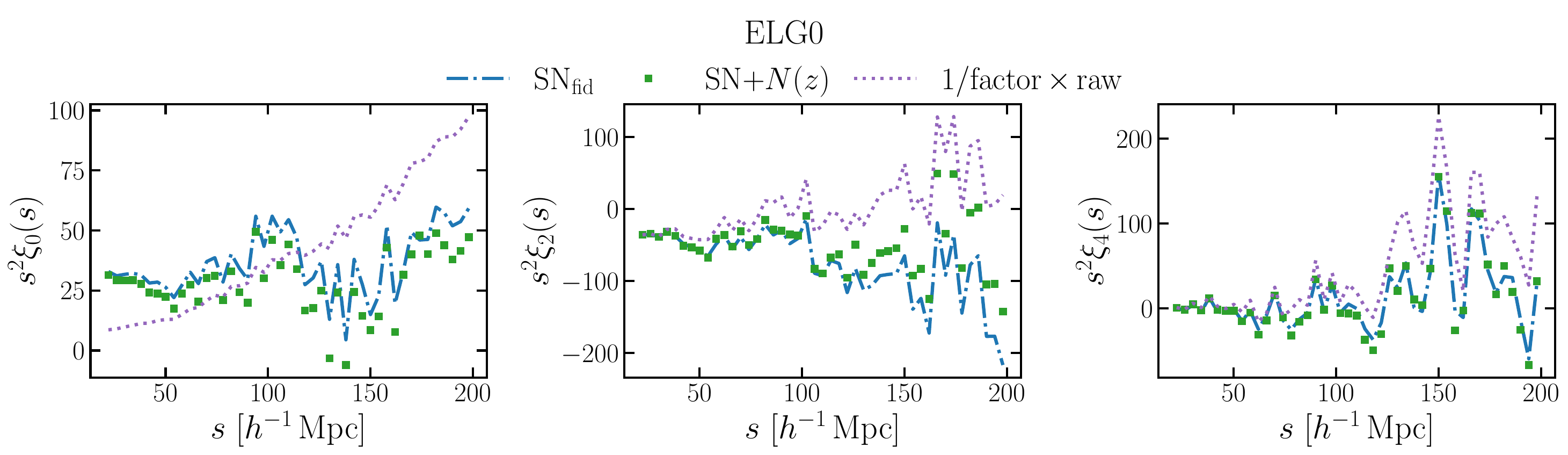}
    \includegraphics[width=\textwidth]{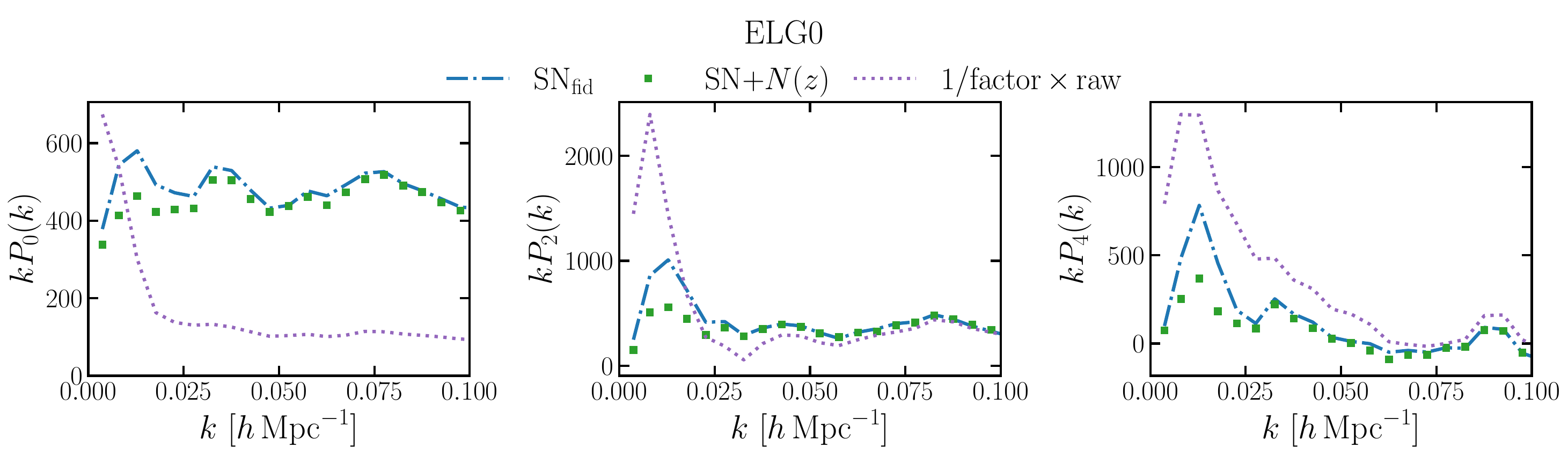}
    \caption{Clustering statistics for the \elgz sample (NGC+SGC). 
    From left to right, the figure shows the monopole, quadrupole, and hexadecapole. The top panels correspond to the \elgz 2-pt correlation function, while the bottom panels show the \elgz power spectra.
    The dotted dashed blue line corresponds to \sysnet weights derived for the ELGs within the $0.6<z<0.8$ redshift range, these weights use the same data preparation method and \sysnet settings employed for the fiducial weights.
    The green squares show data weighted with \sysnet weights derived from training the NN on $N(z)$ weighted data, and then we multiply by $N(z)$ weights. 
    The dotted purple lines show when no angular or radial imaging weights are applied. 
    To facilitate comparison of the \elgz sample, the monopole of the `raw' signal is divided by a factor of 5, and the rest of the multipoles are left alone. }
    \label{fig:dndz_clustering}
\end{figure}

Now that we have discussed the naming scheme for this subsection, let us discuss the clustering statistics that result from applying these weights. In Fig~\ref{fig:dndz_clustering} we present our clustering results in configuration (top panel) and Fourier (bottom panel) space. Notice that in this figure we do not plot fractional errors since no analytical covariances are available for the \elgz sample. Therefore, we opted to show the unblinded galaxy clustering for the DR1 ELGs within $0.6<z<0.8$. In this figure, the line-style color combination is common for both sub-plots.

Starting from the clustering signal weighted by \sysnet only weights ($\mathrm{SN}_\mathrm{fid}$; dotted-dashed blue), we observe that the clustering amplitude across all multipoles in configuration space is consistently lower than the unweighted clustering signal (raw; purple). In Fourier space, the same can be said when we look at low values of $k$, as seen in the previous subsection. 
Then, when we apply the combined \sysnet and depth-dependent $N(z)$ weighting scheme ($\mathrm{SN}+N(z)$; green squares), we find that the power spectrum multipoles amplitude and the correlation function monopole amplitude are consistently lower than that from the $\mathrm{SN}_\mathrm{fid}$ case over relevant scales, i.e. large scales. In the 2-pt correlation function quadrupole a small positive offset occurs around the separation distance of $\sim125\mpch$ when comparing to the $\mathrm{SN}_\mathrm{fid}$ case, while the hexadecapole seems consistent with the same weighting scheme.  

Considering the effect of the N(z) weights, one can observe that the power in all multipoles is significantly higher at $k<0.025$ when the N(z) weight is not included, and that the relative effect is greatest for the quadrupole, which has its amplitude nearly halved. 
The results make it clear that the measured clustering of DESI ELGs with $0.6<z<0.8$ is indeed sensitive to the varying N(z), even without a proper covariance estimate. In addition, the 1D validation of the number density trends with depth also suggest the inclusion of N(z) weights. Hence, we recommend the inclusion of N(z) weights we have defined as well as further study, e.g., simulating the effects on mocks, to attempt to use clustering results from the sample in the future.

\subsection{Impact of imaging systematics on BAO measurement}
\label{subsec:BAO_impact}
To quantify the impact of the imaging systematics and their mitigation schemes on the clustering statistics in terms of the resulting cosmological information, we present the BAO measurements performed on the two-point correlation function with different selection functions, not only for ELGs, but also for all DESI DR1 tracers. 
We performed the BAO fits as described in \cite{optimal_recon}, using \texttt{desilike}\footnote{\hyperlink{https://github.com/cosmodesi/desilike}{https://github.com/cosmodesi/desilike}}.
As the BAO measurement, we constrain the isotropic BAO shift, $\aiso$, and the anisotropic BAO shift, $\aAP$, which are defined as $\aiso\equiv\alpara^{1/3}\alperp^{2/3}$ and  $\aiso\equiv\alpara/\alperp$. More details can be found in \cite{DESI2024.III.KP4}.
Dilation parameters $\alpara$ and $\alperp$ describe the BAO scale along the line of sight and across the line of sight, respectively, and relate to the rate of expansion of the universe and cosmological distances by
\begin{equation}
    \alpha_{\|} = \frac{H^\mathrm{fid}(z) r^\mathrm{tem}_s}{H(z)r_s},       
    \alpha_\perp = \frac{D_\mathrm{A}(z) r^\mathrm{tem}_s}{D^\mathrm{fid}_\mathrm{A}(z)r_s}, 
\end{equation}
where $H(z)$ is the Hubble parameter, $D_\mathrm{A}(z)$ is the angular diameter and $r_s$ is the sound horizon scale at the drag epoch; also `fid' and `tem' are used to represent quantities that follow the fiducial and template cosmologies respectively, as defined in \cite{DESI2024.III.KP4}.
The fits were done using the two-point correlation function  between $50-150\,\mpch$,  
 binned into $4\,\mpch$ separation bins, following the DESI DR1 default method~\cite{KP4s2-Chen, DESI2024.III.KP4}. The default fitting setup for each redshift bin matches the convention adopted by the DESI collaboration. Specifically, only an isotropic BAO fit was done on \bgs, \elgo, and \qso due to its low signal-to-noise, while the 2D fit was performed for the rest.

\begin{figure}
    \centering  
    \includegraphics[width=\textwidth]{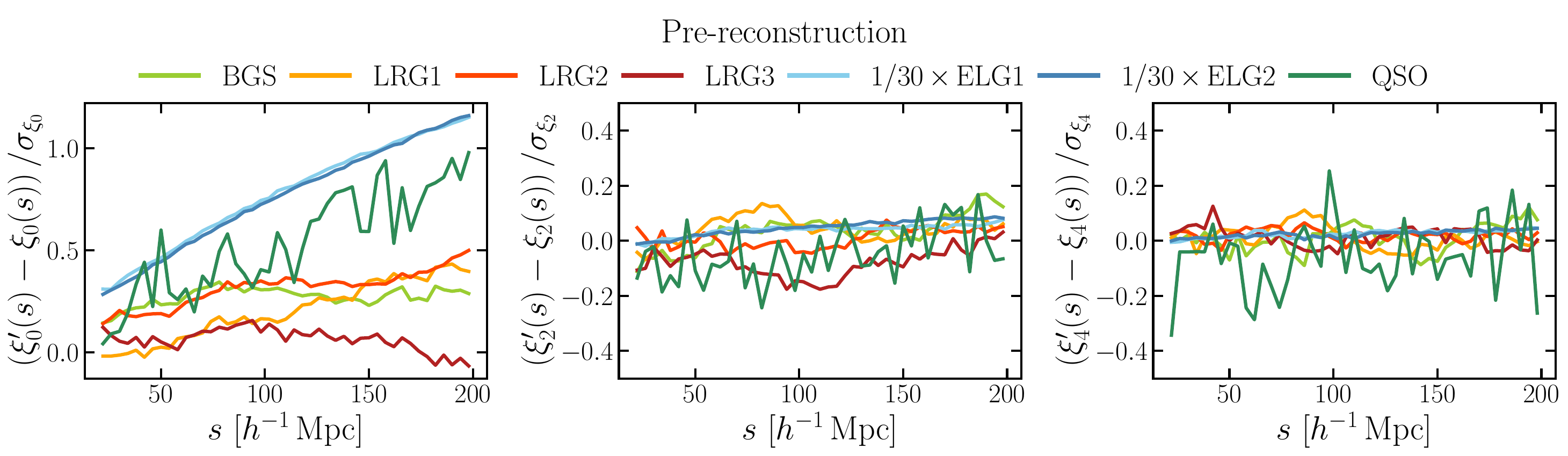}
    \includegraphics[width=\textwidth]{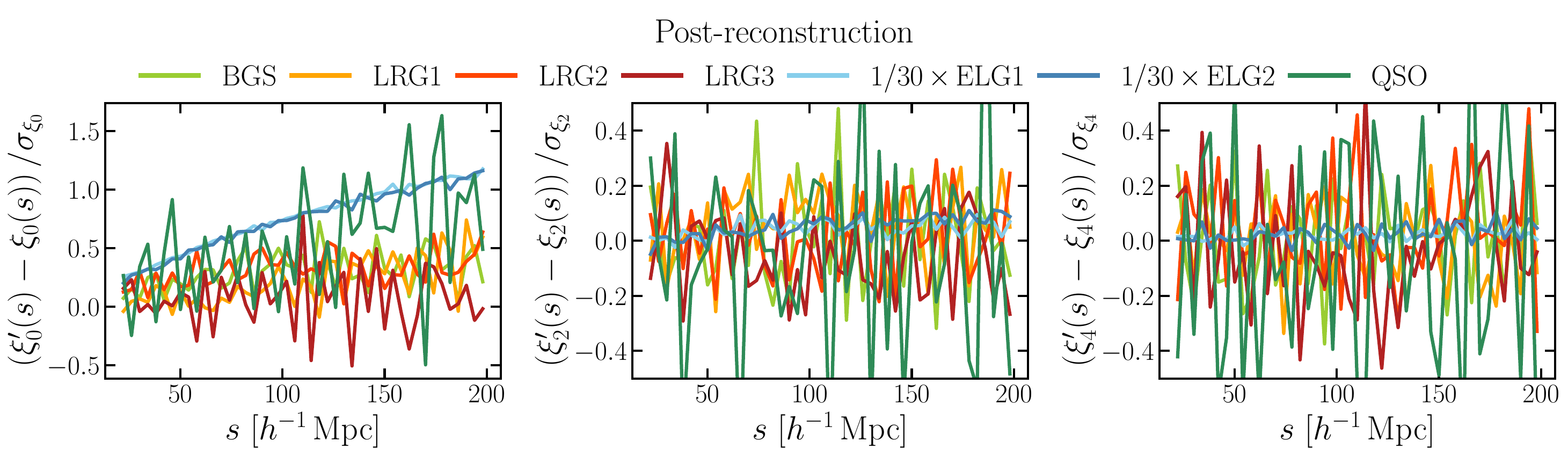}
    \caption{Fractional errors of the differences between the correlation function multipoles for the unweighted $\xi^\prime_\ell(s)$ and weighted $\xi_\ell(s)$ tracers  (NGC+SGC). From left to right, the figure shows the monopole, quadrupole, and hexadecapole. The top and bottom panels correspond to pre- and post-recon results respectively. The ELG results, as expected, are heavily contaminated, hence we divide them by a factor of 30 for convenience.}
    \label{fig:all_xi_sys}
\end{figure}

In Fig.~\ref{fig:all_xi_sys} we show the fractional errors of the 2-pt correlation function multipoles for all DESI tracers, pre- and post-reconstruction\footnote{Reconstruction is performed separately for the weighted and unweighted galaxy density fields.}. The correlation functions shown in this figure are the observables used in the BAO fit results presented in this work. Notice how the ELGs are the most contaminated tracer, showing the highest excess clustering amplitude across all tracers. Considering that these results are dominated by contamination due to imaging systematics, we still do not observe any distinct BAO-like feature in the residuals shown in Fig.~\ref{fig:all_xi_sys}.

\begin{figure}
    \centering
    \includegraphics[width=0.70\textwidth]{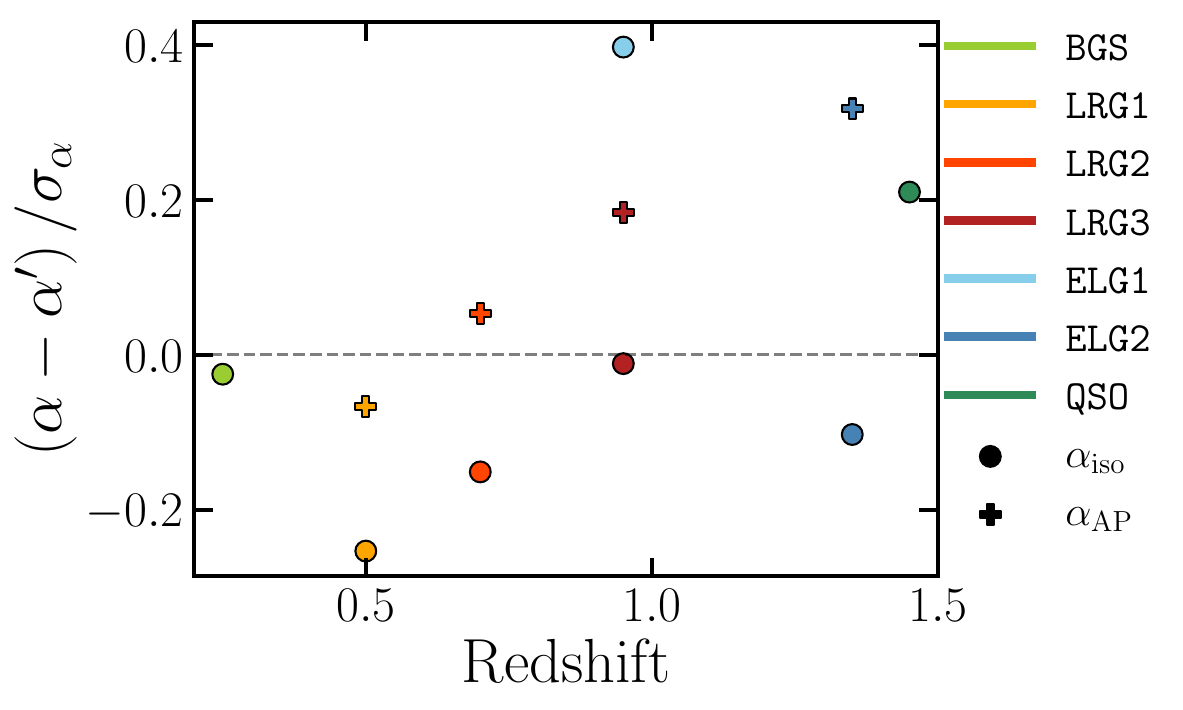}
    \caption{Summary of the BAO measurement results for all DESI tracers (post-reconstruction). On the y axis are the BAO shift fractional errors, where $\alpha$ is obtained using the default imaging weights and $\alpha^\prime$ is from fitting the raw signal. The $\sigma_\alpha$ is the uncertainty for the $\alpha$ result. The circle markers represent the isotropic BAO shift fractional errors; the plus markers represent the anisotropic BAO shift fractional errors.}
    \label{fig:BAO_all}
\end{figure}

Table \ref{tab:alphadiff_table} shows the summary for the isotropic and anisotropic BAO results for the default weight scheme for each tracer: BGS and LRGs with the linear weight, ELGs with \textsc{SYSNet}, and  QSO \textsc{Regressis}, as specified in \cite{DESI2024.II.KP3}. 
Figure~\ref{fig:BAO_all} shows differences between the BAO measurements from the raw clustering statistics (i.e., without any mitigation) and those with the default mitigation. Despite the substantial impact of the imaging systematics and the mitigation observed in terms of the clustering statistics (particularly for ELGs, as shown with the raw clustering signal in Fig.~\ref{fig:xi_sys_ELGS_NS} and \ref{fig:pk_sys_ELGS_NS}), the differences in the BAO measurements are negligible compared to the statistical error and are likely random variations, given that we observe no systematic trend in the direction of the fluctuations. Therefore, we confirm that the imaging systematics do not systematically interfere with the DESI DR1 BAO measurement, as the BAO feature is a characteristic physical scale that is distinguishable from the impact of the imaging systematics on the broadband.

On the other hand, the impact on the full shape measurement~\citep{DESI2024.V.KP5} is more significant and requires a recipe to derive the robust full shape measurements while marginalizing over the effect of potential systematics. Such discussion and methods adopted are presented in \cite{KP5s6-Zhao} and we refer the readers to that work.

\begin{table}
    \centering
    \caption{Summary of the isotropic and anisotropic BAO shift fractional errors results for all DESI tracers (pre- and post-reconstruction). Here the deltas follow the same convention as in Fig~.\ref{fig:BAO_all}, where $\Delta\alpha=(\alpha-\alpha^\prime)$.
    The columns like $x\sigma_{\alpha}$ are the BAO shift fractional errors as presented in Fig~\ref{fig:BAO_all}.}
    \label{tab:alphadiff_table}
    \begin{tabular}{|l|c|cc|cc|}
        \hline
          Tracer   & Recon   &   $\Delta \alpha_\mathrm{iso}$ &   $x\sigma_{\alpha_\mathrm{iso,weighted}}$ & $\Delta \alpha_\mathrm{AP}$   & $x\sigma_{\alpha_\mathrm{AP,weighted}}$   \\
        \hline
         \hline
         \tt{BGS}  & Post    &                        -0.0005 &                       -0.025  & ---                           & ---                           \\
         \tt{LRG1} & Post    &                        -0.003  &                       -0.2534 & -0.0025                       & -0.0668                       \\
         \tt{LRG2} & Post    &                        -0.0017 &                       -0.1512 & 0.0022                        & 0.0535                        \\
         \tt{LRG3} & Post    &                        -0.0001 &                       -0.0114 & 0.0056                        & 0.1837                        \\
         \tt{ELG1} & Post    &                         0.008  &                        0.3974 & ---                           & ---                           \\
         \tt{ELG2} & Post    &                        -0.0016 &                       -0.1029 & 0.0146                        & 0.3175                        \\
         \tt{QSO}  & Post    &                         0.0049 &                        0.2101 & ---                           & ---                           \\
        \hline
         \tt{BGS}  & Pre     &                         0.0004 &                        0.014  & ---                           & ---                           \\
         \tt{LRG1} & Pre     &                        -0.0004 &                       -0.0243 & -0.0005                       & -0.0085                       \\
         \tt{LRG2} & Pre     &                        -0.0001 &                       -0.0058 & 0.0010                        & 0.0128                        \\
         \tt{LRG3} & Pre     &                        -0.0008 &                       -0.059  & -0.0021                       & -0.0445                       \\
         \tt{ELG1} & Pre     &                         0.0458 &                        0.6657 & ---                           & ---                           \\
         \tt{ELG2} & Pre     &                         0.0004 &                        0.0226 & -0.0069                       & -0.1105                       \\
         \tt{QSO}  & Pre     &                         0.0017 &                        0.0842 & ---                           & ---                           \\
        \hline
    \end{tabular}
\end{table}

\clearpage
\section{Conclusion}\label{sec:conclusion}
In this paper, we have investigated angular imaging systematics of DESI DR1 ELGs and developed mitigation methods using NN-based regression (\textsc{SYSNet}) as well as forward modeling (\textsc{Obiwan}). 
We studied the radial systematics of the same sample and have corrected them across the complete DESI ELG redshift range, $0.6<z<1.6$. All mitigation schemes were compared and assessed with $\chi^2$ statistics.
We presented evidence of the robustness of the DESI DR1 BAO measurement across all DESI DR1 tracers in their respective redshift bins. We summarize our key findings:

\begin{itemize}
    \item DESI ELGs are faint in imaging making them highly sensitive to imaging systematics, particularly galactic extinction ($E(B-V)$), galactic depth ($\mathrm{galdepth}_b$), and median seeing ($\mathrm{psfsize}_b$). We were able to make an angular selection function that mitigates these known imaging systematics using the \sysnet pipeline. The imaging weights generated by this method are the fiducial systematic weights for DESI DR1 ELGs used throughout the DESI DR1 cosmological analysis.
    
    \item We find good qualitative agreement between \obiwan predictions of the strength and slope of trends between imaging properties and DESI ELG density. Crucial to this agreement is simulating the effect of the difference in \ebv assumed in targeting and recently measured by DESI \cite{KP3s14-Zhou}. This demonstrates that the causes of the trends in DESI ELG density are well-understood. However, in detail, the statistical agreement is not good and prevents us from directly using the \obiwan to correct for DESI ELG imaging systematics. The lack of agreement is likely due to a combination of residual uncertainty in the DESI \ebv map and imperfections in the \obiwan forward-modeling pipeline.

    \item  We found that ELGs not only suffer from angular systematics
    but also that some imaging properties such as galactic depth in different optical bands cause spurious $N(z)$ variations through target selection, particularly within the $0.6<z<0.8$ redshift range. 
    This variation in $N(z)$ is stronger for the lowest and highest depth quintiles. The lowest depth quintile slightly prefers lower redshift ELGs, while the highest depth quintile slightly prefers higher redshift ELGs, due to the deeper imaging. 
    This variation in the N(z) is strongest for $z < 0.8$ and is the primary reason that the default DESI DR1 LSS catalogs include only ELG data with $0.8<z<1.6$.
    
    \item We found that \obiwan simulations reproduce the observed $N(z)$ variations as a function of galactic depth in the DESI DR1 ELG sample to a good extent. Hence, we used \obiwan to develop a simple model that removes the fluctuations in the observed $N(z)$, especially in the $0.6<z<0.8$ range.
 
    \item We developed and explored a hybrid mitigation method that combines forward modeling and NN-based regression, which performs as well as the sole NN-based approach (DESI DR1 ELGs fiducial mitigation method).
    We found agreement between the mitigation approaches when comparing their clustering statistics, qualitatively through the fractional differences of their multipoles and quantitatively through their $\chi^2$ statistics.
    In this process, we found that the fluctuations \obiwan predicts are captured by the \sysnet mitigation method. 
   
    \item We construct a full three-dimensional selection function through the combination of angular systematic weights obtained with the NN-based regression (\textsc{SYSNet}) and radial systematic weights obtained by modeling the $N(z)$ of simulated ELGs (\textsc{Obiwan}) as a function of galactic depth. Clustering statistics showed that the impact of adding radial weights is negligible for ELGs within $0.8<z<1.6$. Hence, radial weights are not included in fiducial weights since their effect is negligible for the redshifts used for cosmological analysis.
    While for $0.6<z<0.8$ we do find a significant reduction in the clustering amplitude over all power spectrum multipoles.

    \item We perform BAO measurements for all DESI tracers before and after applying their fiducial mitigation schemes, and find that the DESI DR1 BAO measurement is robust against how we mitigate the imaging systematics, even under strong contamination (as is the case for ELGs).

\end{itemize}

While we demonstrate that the effect of the imaging systematics is negligible for the BAO, which is a distinct signature, the mitigation of robust imaging systematics is often accompanied by over-correction of the cosmological clustering and impacts the accurate measurement of broadband features on very large scales, such as primordial non-Gaussianity. Being able to limit such over-correction while effectively removing systematics is a critical improvement we need to make \cite{rezaie2023local}. In this paper, we explored the idea of a hybrid approach, which has the potential to limit the unnecessary flexibility in the mitigation and therefore restrain the over-correction effect. 
For future work, it is ideal to find the minimum set of templates that can be used in the second layer of training \sysnet (Section~\ref{subsec:second_step}) such that we can significantly ameliorate systematics while minimizing the amount of over-correction from giving \sysnet redundant information. 
A possible way to achieve this is by using \obiwan results as an input map, then the combined $\obiwan+\sysnet$ approach would account for the effort of the forward simulation, ideally allowing for a reduced set of input imaging maps to be given to the NN.

\section*{Acknowledgements}
AJRM and H-JS acknowledge support from the U.S. Department of Energy, Office of Science, Office of High Energy Physics under grant No. DE-SC0023241 and DE-SC0019091.

This material is based upon work supported by the U.S. Department of Energy (DOE), Office of Science, Office of High-Energy Physics, under Contract No. DE–AC02–05CH11231, and by the National Energy Research Scientific Computing Center, a DOE Office of Science User Facility under the same contract. Additional support for DESI was provided by the U.S. National Science Foundation (NSF), Division of Astronomical Sciences under Contract No. AST-0950945 to the NSF’s National Optical-Infrared Astronomy Research Laboratory; the Science and Technology Facilities Council of the United Kingdom; the Gordon and Betty Moore Foundation; the Heising-Simons Foundation; the French Alternative Energies and Atomic Energy Commission (CEA); the National Council of Humanities, Science and Technology of Mexico (CONAHCYT); the Ministry of Science, Innovation and Universities of Spain (MICIU/AEI/10.13039/501100011033), and by the DESI Member Institutions: \url{https://www.desi.lbl.gov/collaborating-institutions}.

The DESI Legacy Imaging Surveys consist of three individual and complementary projects: the Dark Energy Camera Legacy Survey (DECaLS), the Beijing-Arizona Sky Survey (BASS), and the Mayall z-band Legacy Survey (MzLS). DECaLS, BASS and MzLS together include data obtained, respectively, at the Blanco telescope, Cerro Tololo Inter-American Observatory, NSF’s NOIRLab; the Bok telescope, Steward Observatory, University of Arizona; and the Mayall telescope, Kitt Peak National Observatory, NOIRLab. NOIRLab is operated by the Association of Universities for Research in Astronomy (AURA) under a cooperative agreement with the National Science Foundation. Pipeline processing and analyses of the data were supported by NOIRLab and the Lawrence Berkeley National Laboratory. Legacy Surveys also uses data products from the Near-Earth Object Wide-field Infrared Survey Explorer (NEOWISE), a project of the Jet Propulsion Laboratory/California Institute of Technology, funded by the National Aeronautics and Space Administration. Legacy Surveys was supported by: the Director, Office of Science, Office of High Energy Physics of the U.S. Department of Energy; the National Energy Research Scientific Computing Center, a DOE Office of Science User Facility; the U.S. National Science Foundation, Division of Astronomical Sciences; the National Astronomical Observatories of China, the Chinese Academy of Sciences and the Chinese National Natural Science Foundation. LBNL is managed by the Regents of the University of California under contract to the U.S. Department of Energy. The complete acknowledgments can be found at \url{https://www.legacysurvey.org/}.

Any opinions, findings, and conclusions or recommendations expressed in this material are those of the author(s) and do not necessarily reflect the views of the U. S. National Science Foundation, the U. S. Department of Energy, or any of the listed funding agencies.

The authors are honored to be permitted to conduct scientific research on Iolkam Du’ag (Kitt Peak), a mountain with particular significance to the Tohono O’odham Nation.

\section*{Data Availability}
\label{sec:dataavail}

The \obiwan ELGs catalog is available internally for DESI members\footnote{\url{https://data.desi.lbl.gov/desi/survey/catalogs/image_simulations/ELG/}}. 
The data will be uploaded to Zenodo before publication\footnote{\url{https://zenodo.org/uploads/}}.

% references
\bibliographystyle{JHEP}
\bibliography{refs,DESI2024_updated15Aug} 

\appendix

\section{\sysnet settings}\label{sec:SN_settings}
For NN training, the hyper-parameters for \elgo North and \elgt North are as follows: we use a NN structure of 10 neurons with 3 hidden layers, with a batch size $N_\mathrm{batch}=256$ and a learning rate $\lambda=0.009$. For \elgo South and \elgt South, we use a NN structure of 20 neurons with 4 hidden layers, with a batch size $N_\mathrm{batch}=1024$ and a learning rate $\lambda=0.007$. 
In addition, all samples were trained for 100 epochs, using five-fold cross-validation with $N_\mathrm{chains}=5$, which results in an ensemble of 25 NN models for each sample. In Table \ref{tab:sn_hyperparameters} we show a summary of the hyper-parameters and options we used in the NN training for each sample.

\begin{table}
    \centering
    \begin{tabular}{ccccccccc}
    \hline
         & Tracer & Region & Redshift & $\mathrm{NN}_\mathrm{structure}$ & $\lambda$ & $N_\mathrm{batch}$ & $N_\mathrm{epochs}$ & $N_\mathrm{chains}$ \\
         \hline
         \hline
         & ELG1 & $\mathbf{North}$ & $0.8<z<1.1$ & $(3,10)$ & $0.009$ & $256$ & 100 & 5 \\
         & ELG2 &                  & $1.1<z<1.6$ & $(3,10)$ & $0.009$ & $256$ & 100 & 5 \\
         \hline
         & ELG1 & $\mathbf{South}$ & $0.8<z<1.1$ & $(4,20)$ & $0.007$ & $1024$ & 100 & 5 \\
         & ELG2 &                  & $1.1<z<1.6$ & $(4,20)$ & $0.007$ & $1024$ & 100 & 5 \\
         \hline
    \end{tabular}
    \caption{Hyper-parameters used for \sysnet training on DESI ELGs. Both regions were further split into two redshift bins ($0.8<z<1.1$ and $1.1<z<1.6$) before training, and we set each run for 100 epochs, number of chains was set to 5 and $k_\mathrm{fold}=5$ cross-validation was used. In $\mathrm{NN}_\mathrm{structure}$ the first integer indicates the number of hidden layers, the second is the number of neurons in each layer.}
    \label{tab:sn_hyperparameters}
\end{table}

\begin{figure}
    \centering
    \includegraphics[width=0.8\textwidth]{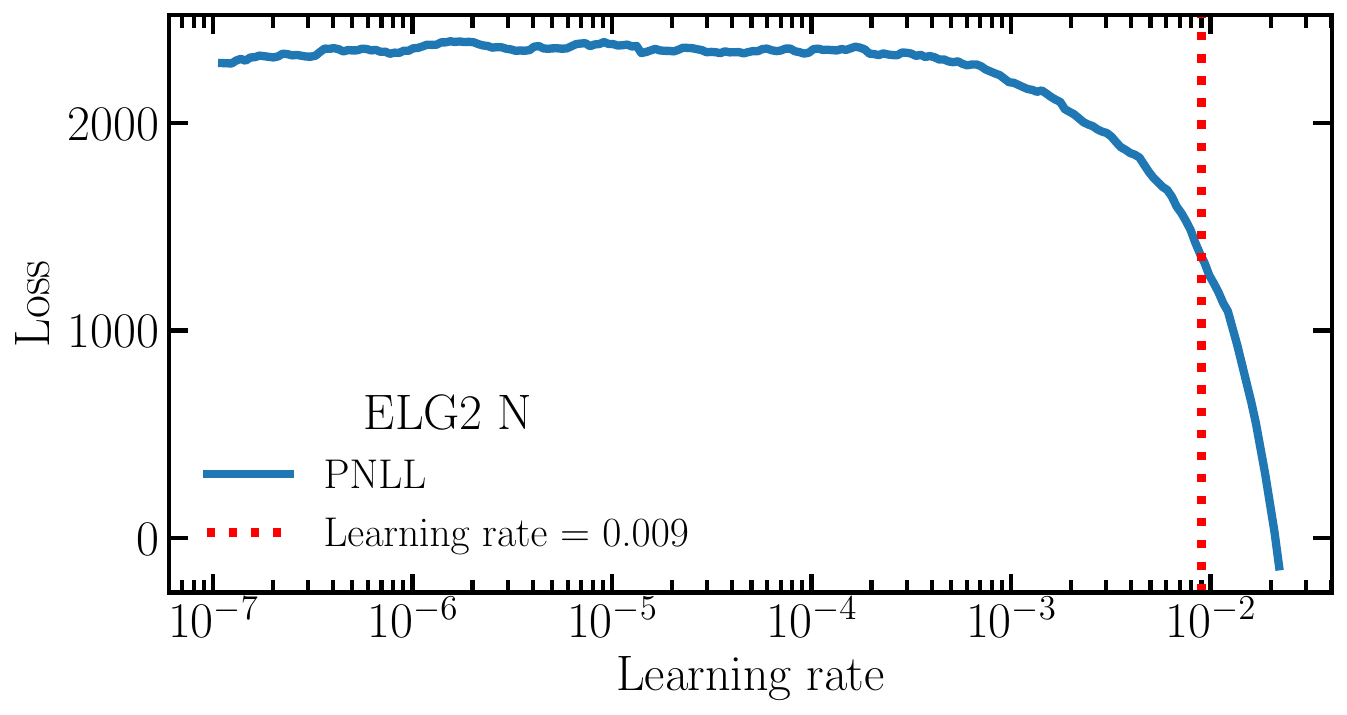}
    \caption{Loss as a function of learning rate for the \elgt North sample, with this LRRT we choose an optimal learning rate of 0.009 (dotted solid red line), the optimal learning rate is chosen at a point where the cost function vs learning rate is steeply decreasing. In this plot, PNLL means Poisson negative log-likelihood.}
    \label{fig:loss_vs_lr}
\end{figure}

To determine the adopted learning rates described above, we ran the NN for 1 epoch leaving the learning rate to fluctuate within some boundary and then selected the learning rate $\lambda$ that allowed for the steepest negative gradient. This is referred to as a learning rate range test (LRRT) \cite{smith2017cyclicallearningratestraining}. In Fig. \ref{fig:loss_vs_lr} we show an example of how the optimal $\lambda$ was chosen for the \elgt North sample, the same procedure was applied to the other samples. The process is not automated. Instead, we inspected the results shown in Fig. \ref{fig:loss_vs_lr} (a similar figure is produced for each case) and chose the value of $\lambda$, by-eye, where the slope is most negative within a region where the curve is smooth.

\begin{figure}
    \centering
    \includegraphics[width=\textwidth]{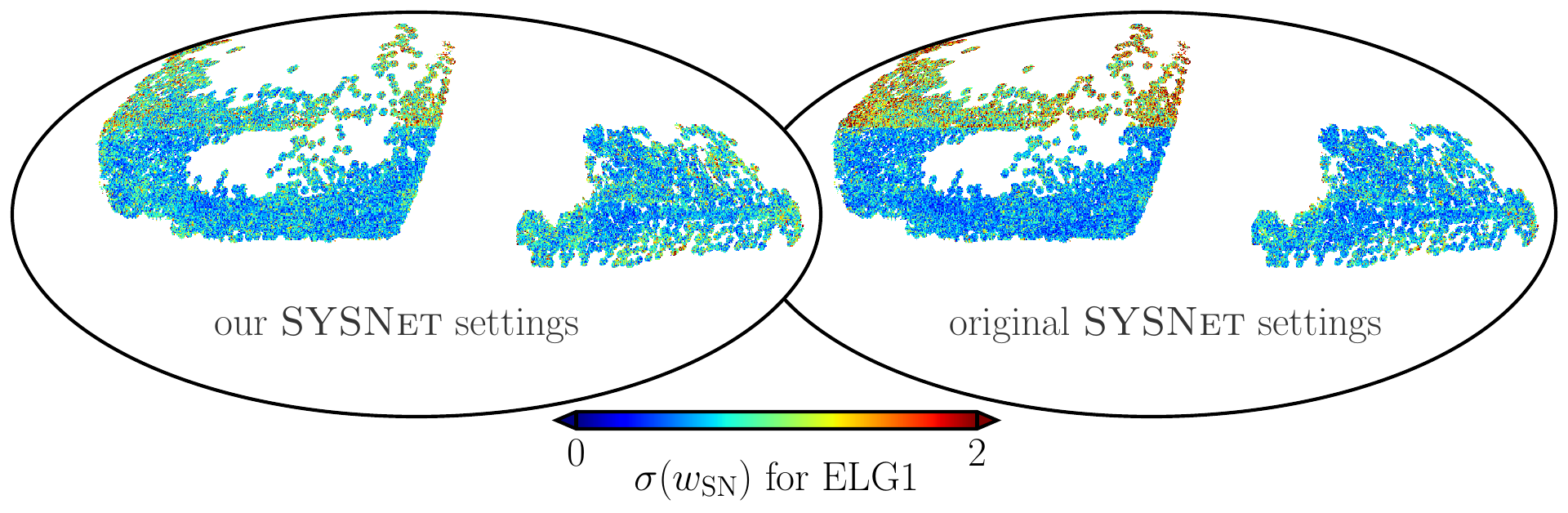}
    \includegraphics[width=\textwidth]{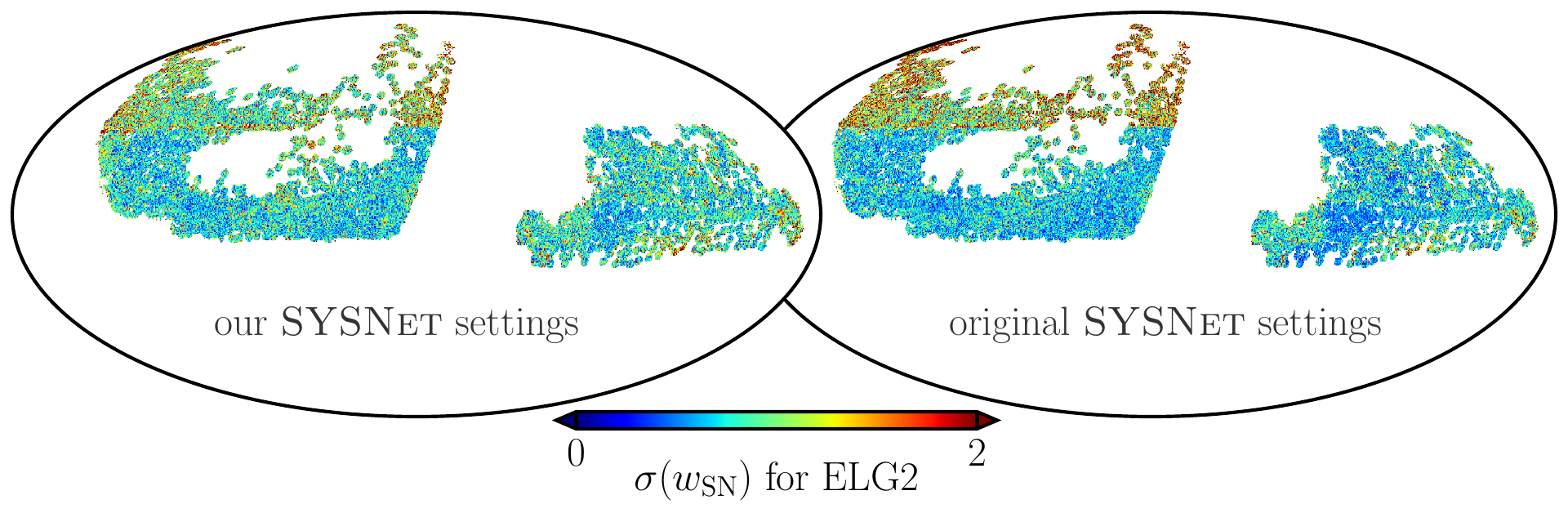}
    \caption{Dispersion across the 25 NN predictions at the pixel level. The left panels show the dispersion when using our choice of hyper-parameters, while the right panels show the dispersion for the original choice of settings. The top and bottom show results for the \elgo and \elgt respectively. Qualitatively, our choice of \sysnet settings improves the stability of NN predictions for the North, while they remain similar for the South.}
    \label{fig:sigma_SN}
\end{figure}

The settings we use for batch-size and NN structure differ slightly from the original values found in the \sysnet pipeline configuration file. The original values for $N_\mathrm{batch}$ and $\mathrm{NN}_\mathrm{structure}$ are 512 and (4,20) respectively. However, at the suggestion of Mehdi Rezaie (developer of \textsc{SYSNet}), we use the settings presented in Table~\ref{tab:sn_hyperparameters}.  Notice that for the North samples, we use a simpler NN model (3,10) since the sample contains $\sim3.7$ times fewer pixels than the South, and we reduce the batch-size to 256. For the South, to improve efficiency (less computing time) we increase the batch-size to 1024. In Fig.~\ref{fig:sigma_SN} we show skymaps of the dispersion across the 25 NN predictions at the pixel level, we call this dispersion $\sigma(w_\mathrm{SN})$. Qualitatively, from Fig.~\ref{fig:sigma_SN}, we find that our choice of settings improves the stability of the NN predictions in the North; while for the South we maintain similar stability between the choice of settings, with the benefit of more efficiency.

\section{Implementation details for producing the \obiwan ELG sample}
\label{sec:obiwan_appendix}
\begin{figure}
    \centering  
    \includegraphics[width=0.7\textwidth]{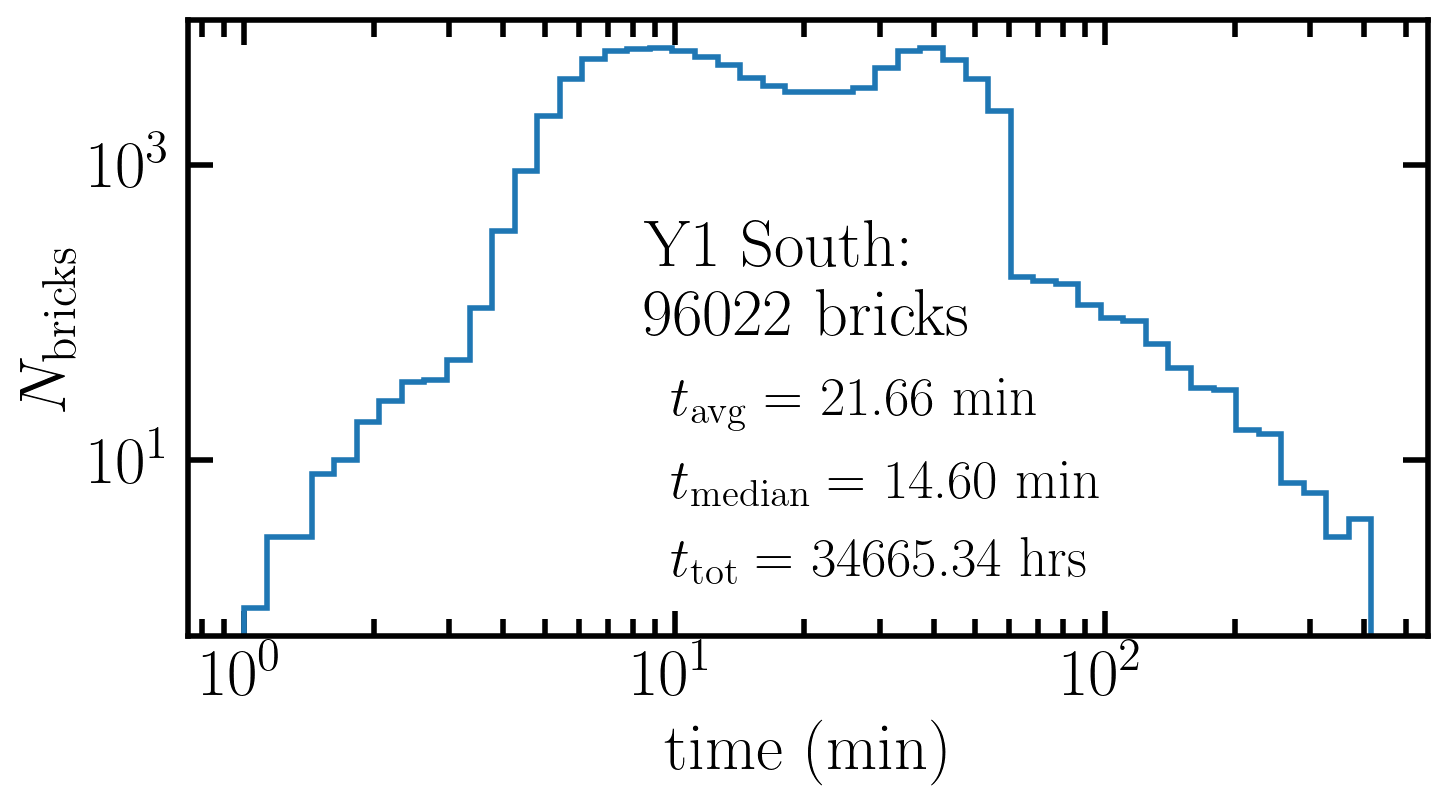}
    \caption{\obiwan runtime for bricks in the DESI DR1 South sample (excluding bricks from the M2 South sample). The median (average) time to process one brick is 14.60 (21.66) minutes.
    Each brick was completed using 8 cores (16 threads) on 1 NERSC perlmutter node.}
    \label{fig:obi_runtime}
\end{figure}

Both samples (M2 and the remainder of DR1) were completed on separate machines, computing times vary significantly. Hence, we show computing times for M2 and DR1 in the North and South separately. The amount of bricks that can be run in parallel per node is limited by memory availability, we have found that allowing for 32GB per brick is generally enough to avoid the process running out of memory. Since the Cori machine has (no longer available) 128GB of available memory per node, we consistently ran M2 with 4 bricks per computing node; while Perlmutter has 512GB of available memory per node, we ran the remaining of DR1 with 16 bricks per computing node. Given these details, now we present the median and average run times per brick for each sample, with quantities in parenthesis divided by the number of bricks run in parallel: M2 North median time 39.88 (9.97) mins and average 43.34 (10.84) mins, M2 South median time 58 (14.5) mins and average 71.35 (17.84) mins, DR1 North median time 28.16 (1.76) mins and average 33.16 (2.07) mins, and DR1 South median time 14.60 (0.91) mins and average 21.66 (1.35) mins.
In Fig~\ref{fig:obi_runtime} we show, as an example, the run time for DR1 South sample, notice that the total of $\sim$34,665 hours is not the real node hours used to complete the sample since we run 16 bricks per node, therefore a real estimate of the total node hours used would be $\sim$2,166 hours.

%\input{affiliations}
% Don't change these lines
%\label{lastpage}
\end{document}

%% file: authorlist.tex
\emailAdd{ar652820@ohio.edu}

\author[1]{{A.~J.~Rosado-Mar\'{i}n}\orcidlink{0000-0001-7545-3504},}
\author[2,3,4]{{A.~J.~Ross}\orcidlink{0000-0002-7522-9083},}
\author[1]{{H.~Seo}\orcidlink{0000-0002-6588-3508},}
\author[5]{{M.~Rezaie}\orcidlink{0000-0001-5589-7116},}
\author[6,4]{{H.~Kong},}
\author[7]{{A.~de~Mattia}\orcidlink{0000-0003-0920-2947},}
\author[8]{{R.~Zhou}\orcidlink{0000-0001-5381-4372},}
\author[8]{{J.~Aguilar},}
\author[9]{{S.~Ahlen}\orcidlink{0000-0001-6098-7247},}
\author[10]{{O.~Alves},}
\author[11]{{D.~Bianchi}\orcidlink{0000-0001-9712-0006},}
\author[12]{{D.~Brooks},}
\author[7]{{E.~Burtin},}
\author[8]{{E.~Chaussidon}\orcidlink{0000-0001-8996-4874},}
\author[13]{{X.~Chen}\orcidlink{0000-0003-3456-0957},}
\author[8]{{T.~Claybaugh},}
\author[14]{{K.S.~Dawson}\orcidlink{0000-0002-0553-3805},}
\author[15]{{A.~de la Macorra}\orcidlink{0000-0002-1769-1640},}
\author[16]{{Arjun~Dey}\orcidlink{0000-0002-4928-4003},}
\author[12]{{P.~Doel},}
\author[17,18]{{K.~Fanning}\orcidlink{0000-0003-2371-3356},}
\author[8,19]{{S.~Ferraro}\orcidlink{0000-0003-4992-7854},}
\author[20,21]{{J.~E.~Forero-Romero}\orcidlink{0000-0002-2890-3725},}
\author[22,23,24]{{E.~Gazta\~naga},}
\author[8]{{S.~Gontcho A Gontcho}\orcidlink{0000-0003-3142-233X},}
\author[25]{{G.~Gutierrez},}
\author[26,27]{{C.~Hahn}\orcidlink{0000-0003-1197-0902},}
\author[10]{{M.~M.~S~Hanif}\orcidlink{0009-0006-2583-5006},}
\author[28]{{C.~Howlett}\orcidlink{0000-0002-1081-9410},}
\author[16]{{S.~Juneau}\orcidlink{0000-0002-0000-2394},}
\author[29]{{R.~Kehoe},}
\author[8]{{T.~Kisner}\orcidlink{0000-0003-3510-7134},}
\author[8]{{A.~Kremin}\orcidlink{0000-0001-6356-7424},}
\author[30,31,32]{{A.~Krolewski},}
\author[8]{{M.~Landriau}\orcidlink{0000-0003-1838-8528},}
\author[33]{{L.~Le~Guillou}\orcidlink{0000-0001-7178-8868},}
\author[8]{{M.~E.~Levi}\orcidlink{0000-0003-1887-1018},}
\author[16]{{A.~Meisner}\orcidlink{0000-0002-1125-7384},}
\author[34]{{J.~Mena-Fern\'andez}\orcidlink{0000-0001-9497-7266},}
\author[35,6]{{R.~Miquel},}
\author[36]{{J.~Moustakas}\orcidlink{0000-0002-2733-4559},}
\author[37]{{J.~ A.~Newman}\orcidlink{0000-0001-8684-2222},}
\author[30,26,32]{{E.~Paillas}\orcidlink{0000-0002-4637-2868},}
\author[7,8]{{N.~Palanque-Delabrouille}\orcidlink{0000-0003-3188-784X},}
\author[30,31,32]{{W.~J.~Percival}\orcidlink{0000-0002-0644-5727},}
\author[38]{{F.~Prada}\orcidlink{0000-0001-7145-8674},}
\author[39]{{I.~P\'erez-R\`afols}\orcidlink{0000-0001-6979-0125},}
\author[8]{{A.~Raichoor}\orcidlink{0000-0001-5999-7923},}
\author[40]{{G.~Rossi},}
\author[41,28]{{R.~Ruggeri}\orcidlink{0000-0002-0394-0896},}
\author[42]{{E.~Sanchez}\orcidlink{0000-0002-9646-8198},}
\author[43]{{E.~F.~Schlafly}\orcidlink{0000-0002-3569-7421},}
\author[8]{{D.~Schlegel},}
\author[44,10]{{M.~Schubnell},}
\author[16]{{D.~Sprayberry},}
\author[15]{{M.~Vargas-Maga\~na}\orcidlink{0000-0003-3841-1836},}
\author[16]{{B.~A.~Weaver},}
\author[45]{{J.~Yu}\orcidlink{0009-0001-7217-8006},}
\author[46]{{H.~Zou}\orcidlink{0000-0002-6684-3997}}

\affiliation[1]{Department of Physics \& Astronomy, Ohio University, 139 University Terrace, Athens, OH 45701, USA}
\affiliation[2]{Center for Cosmology and AstroParticle Physics, The Ohio State University, 191 West Woodruff Avenue, Columbus, OH 43210, USA}
\affiliation[3]{Department of Astronomy, The Ohio State University, 4055 McPherson Laboratory, 140 W 18th Avenue, Columbus, OH 43210, USA}
\affiliation[4]{The Ohio State University, Columbus, 43210 OH, USA}
\affiliation[5]{Department of Physics, Kansas State University, 116 Cardwell Hall, Manhattan, KS 66506, USA}
\affiliation[6]{Institut de F\'{i}sica dâ€™Altes Energies (IFAE), The Barcelona Institute of Science and Technology, Edifici Cn, Campus UAB, 08193, Bellaterra (Barcelona), Spain}
\affiliation[7]{IRFU, CEA, Universit\'{e} Paris-Saclay, F-91191 Gif-sur-Yvette, France}
\affiliation[8]{Lawrence Berkeley National Laboratory, 1 Cyclotron Road, Berkeley, CA 94720, USA}
\affiliation[9]{Physics Dept., Boston University, 590 Commonwealth Avenue, Boston, MA 02215, USA}
\affiliation[10]{University of Michigan, 500 S. State Street, Ann Arbor, MI 48109, USA}
\affiliation[11]{Dipartimento di Fisica ``Aldo Pontremoli'', Universit\`a degli Studi di Milano, Via Celoria 16, I-20133 Milano, Italy}
\affiliation[12]{Department of Physics \& Astronomy, University College London, Gower Street, London, WC1E 6BT, UK}
\affiliation[13]{Physics Department, Yale University, P.O. Box 208120, New Haven, CT 06511, USA}
\affiliation[14]{Department of Physics and Astronomy, The University of Utah, 115 South 1400 East, Salt Lake City, UT 84112, USA}
\affiliation[15]{Instituto de F\'{\i}sica, Universidad Nacional Aut\'{o}noma de M\'{e}xico,  Circuito de la Investigaci\'{o}n Cient\'{\i}fica, Ciudad Universitaria, Cd. de M\'{e}xico  C.~P.~04510,  M\'{e}xico}
\affiliation[16]{NSF NOIRLab, 950 N. Cherry Ave., Tucson, AZ 85719, USA}
\affiliation[17]{Kavli Institute for Particle Astrophysics and Cosmology, Stanford University, Menlo Park, CA 94305, USA}
\affiliation[18]{SLAC National Accelerator Laboratory, 2575 Sand Hill Road, Menlo Park, CA 94025, USA}
\affiliation[19]{University of California, Berkeley, 110 Sproul Hall \#5800 Berkeley, CA 94720, USA}
\affiliation[20]{Departamento de F\'isica, Universidad de los Andes, Cra. 1 No. 18A-10, Edificio Ip, CP 111711, Bogot\'a, Colombia}
\affiliation[21]{Observatorio Astron\'omico, Universidad de los Andes, Cra. 1 No. 18A-10, Edificio H, CP 111711 Bogot\'a, Colombia}
\affiliation[22]{Institut d'Estudis Espacials de Catalunya (IEEC), c/ Esteve Terradas 1, Edifici RDIT, Campus PMT-UPC, 08860 Castelldefels, Spain}
\affiliation[23]{Institute of Cosmology and Gravitation, University of Portsmouth, Dennis Sciama Building, Portsmouth, PO1 3FX, UK}
\affiliation[24]{Institute of Space Sciences, ICE-CSIC, Campus UAB, Carrer de Can Magrans s/n, 08913 Bellaterra, Barcelona, Spain}
\affiliation[25]{Fermi National Accelerator Laboratory, PO Box 500, Batavia, IL 60510, USA}
\affiliation[26]{Steward Observatory, University of Arizona, 933 N, Cherry Ave, Tucson, AZ 85721, USA}
\affiliation[27]{Steward Observatory, University of Arizona, 933 N. Cherry Avenue, Tucson, AZ 85721, USA}
\affiliation[28]{School of Mathematics and Physics, University of Queensland, Brisbane, QLD 4072, Australia}
\affiliation[29]{Department of Physics, Southern Methodist University, 3215 Daniel Avenue, Dallas, TX 75275, USA}
\affiliation[30]{Department of Physics and Astronomy, University of Waterloo, 200 University Ave W, Waterloo, ON N2L 3G1, Canada}
\affiliation[31]{Perimeter Institute for Theoretical Physics, 31 Caroline St. North, Waterloo, ON N2L 2Y5, Canada}
\affiliation[32]{Waterloo Centre for Astrophysics, University of Waterloo, 200 University Ave W, Waterloo, ON N2L 3G1, Canada}
\affiliation[33]{Sorbonne Universit\'{e}, CNRS/IN2P3, Laboratoire de Physique Nucl\'{e}aire et de Hautes Energies (LPNHE), FR-75005 Paris, France}
\affiliation[34]{Laboratoire de Physique Subatomique et de Cosmologie, 53 Avenue des Martyrs, 38000 Grenoble, France}
\affiliation[35]{Instituci\'{o} Catalana de Recerca i Estudis Avan\c{c}ats, Passeig de Llu\'{\i}s Companys, 23, 08010 Barcelona, Spain}
\affiliation[36]{Department of Physics and Astronomy, Siena College, 515 Loudon Road, Loudonville, NY 12211, USA}
\affiliation[37]{Department of Physics \& Astronomy and Pittsburgh Particle Physics, Astrophysics, and Cosmology Center (PITT PACC), University of Pittsburgh, 3941 O'Hara Street, Pittsburgh, PA 15260, USA}
\affiliation[38]{Instituto de Astrof\'{i}sica de Andaluc\'{i}a (CSIC), Glorieta de la Astronom\'{i}a, s/n, E-18008 Granada, Spain}
\affiliation[39]{Departament de F\'isica, EEBE, Universitat Polit\`ecnica de Catalunya, c/Eduard Maristany 10, 08930 Barcelona, Spain}
\affiliation[40]{Department of Physics and Astronomy, Sejong University, 209 Neungdong-ro, Gwangjin-gu, Seoul 05006, Republic of Korea}
\affiliation[41]{Centre for Astrophysics \& Supercomputing, Swinburne University of Technology, P.O. Box 218, Hawthorn, VIC 3122, Australia}
\affiliation[42]{CIEMAT, Avenida Complutense 40, E-28040 Madrid, Spain}
\affiliation[43]{Space Telescope Science Institute, 3700 San Martin Drive, Baltimore, MD 21218, USA}
\affiliation[44]{Department of Physics, University of Michigan, 450 Church Street, Ann Arbor, MI 48109, USA}
\affiliation[45]{Institute of Physics, Laboratory of Astrophysics, \'{E}cole Polytechnique F\'{e}d\'{e}rale de Lausanne (EPFL), Observatoire de Sauverny, Chemin Pegasi 51, CH-1290 Versoix, Switzerland}
\affiliation[46]{National Astronomical Observatories, Chinese Academy of Sciences, A20 Datun Rd., Chaoyang District, Beijing, 100012, P.R. China}